\documentclass[twocolumn,prl,nobibnotes,altaffilletter,amsmath,amssymb,amsfonts]{revtex4-1}
\usepackage[latin1]{inputenc}
\usepackage{graphicx}
\usepackage{amsmath}
\usepackage{latexsym}
\usepackage{epsfig}
\usepackage[caption=false]{subfig}
\begin{document}

\title{Multiple higher-order singularities and iso-dynamics in a simple glass-former model}

\author{Nicoletta Gnan$^1$, Gayatri Das$^2$, Matthias Sperl$^3$, Francesco Sciortino$^{2}$ and Emanuela Zaccarelli$^{1,2}$ }
%\email{YYY}
\affiliation{ $^1$ CNR-ISC Uos Sapienza, Piazzale A. Moro 2, I-00185, Roma, Italy }
\affiliation{ $^2$ Dipartimento di Fisica, Sapienza Universit\`{a} di Roma, Piazzale A. Moro 2, I-00185, Roma, Italy }
\affiliation{ $^3$ Institut f\"{u}r Materialphysik im Weltraum, Deutsches Zentrum f\"{u}r Luft-und Raumfahrt, 51170 K\"{o}ln, Germany }
\date{\today}
\begin{abstract}
We investigate the slow dynamics of a colloidal model with two repulsive length scales, whose interaction potential is the sum of a hard-core  and a square shoulder. Despite the simplicity of the interactions,
Mode-Coupling theory predicts a complex dynamic scenario: a fluid-glass line with two reentrances and a glass-glass line ending with multiple higher-order ($A_3$ or  $A_4$) singularities. In this work we verify the existence of the two $A_4$ points by numerical simulations, observing subdiffusive behaviour of the mean-square displacement and logarithmic decay of the density correlators. Surprisingly, we also discover a novel dynamic behaviour generated by the competition between the two higher-order singularities. This results in the presence of special loci along which the dynamics is identical \textit{at all} length and time scales.
\end{abstract}

\maketitle

{\it Introduction:} Colloidal solutions often display a rich variety of dynamical behaviour. % ,  wider than what has been observed in atomic or molecular systems.  % even if the interactions among colloids are  simple.
Experiments have shown that the addition of a short-range attraction to the excluded volume interaction generates  multiple dynamically arrested (glassy) states, explored by  tuning the interaction strength, the packing fraction or  the range of the attraction\cite{Mal00a,Eck02a,Pha02a,Chen03a,Pha04a,Sprung08}.  These experimental studies have confirmed predictions based on the Mode Coupling Theory (MCT)~\cite{Gotzebook,Fab99a,Ber99a,Daw00a,Gotzesperl} and numerical simulations~\cite{Pue02a,Fof02a,Zac02a,Pue03a,Sci03a,Zac04b}.  Unconventional dynamics have also been reported for purely repulsive colloidal mixtures. In particular, multiple glasses have been found in binary mixtures of hard \cite{imhof95,Voigt2011} and soft spheres\cite{Moreno06PRE,Moreno06JCP,voigtmannhorbach}. Also in star-polymer mixtures, distinct glasses have been found to surround a region of ergodic state points~\cite{MayerNature,MayerMacro}.  In all these cases, different microscopic mechanisms compete to generate multiple arrested states.

\begin{figure}
%\begin{center}
\includegraphics[width=0.48\textwidth]{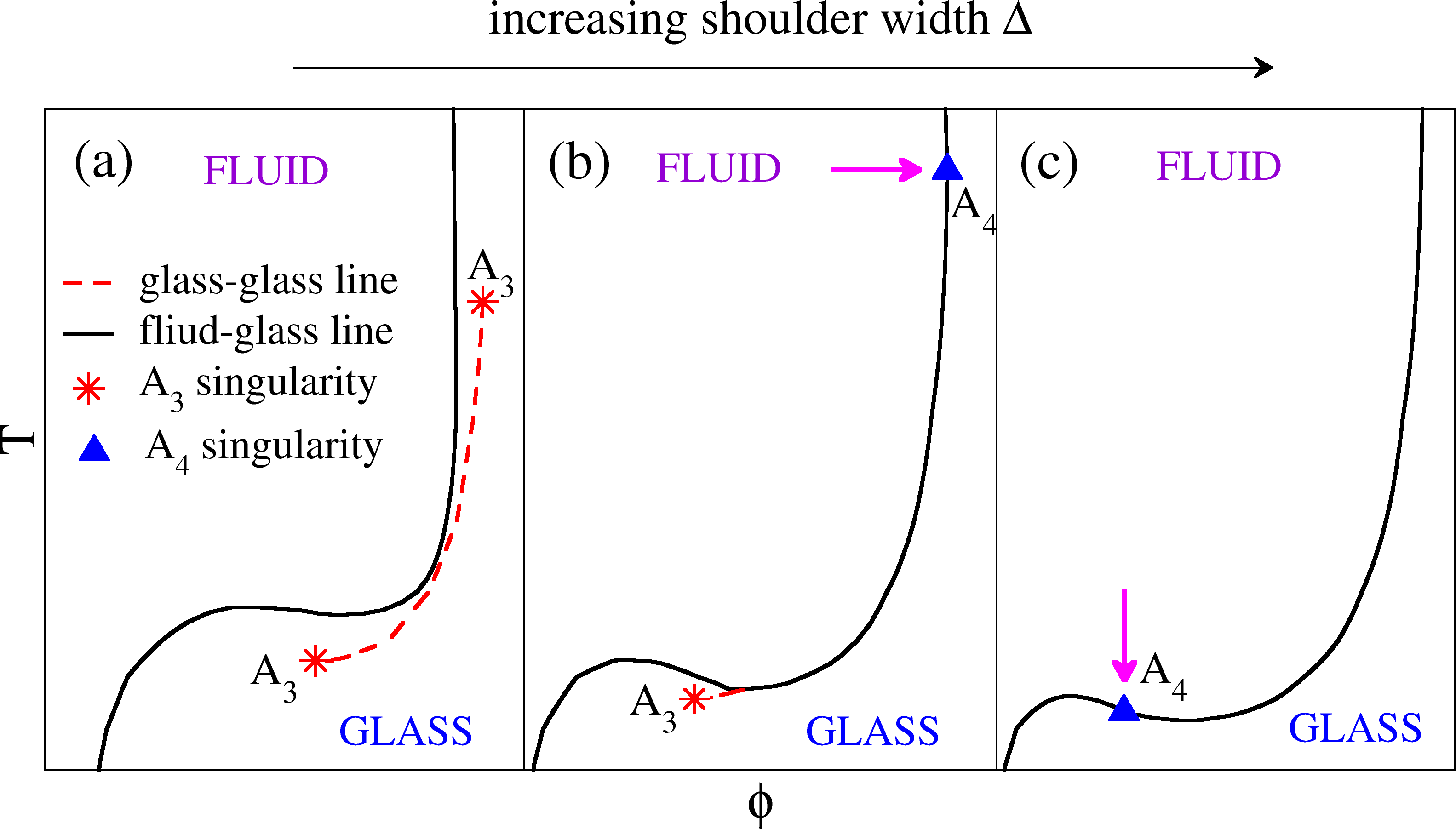}
\caption{
Schematic evolution of the MCT dynamic state diagram of the SS system for increasing values of the shoulder width $\Delta$. (a) For small $\Delta$, beside the fluid-glass line (solid line), a disconnected glass-glass line is predicted (dashed line),  ending in two $A_3$ higher-order singularities (stars). (b) On increasing $\Delta$, the glass-glass line merges with the fluid-glass line and an $A_4$ singularity appears when one of the two $A_3$ points meets the fluid-glass line (filled triangle). (c) At even larger $\Delta$ also the second $A_3$ point eventually intersects the fluid-glass line  generating a distinct $A_4$ higher-order singularity (filled triangle). The arrows in (b) and (c) indicate the paths followed to locate the $A_4$ singularities in the simulations.}
\label{fig:MCTlines}
%\end{center}
\end{figure}

Recently,  core softened potentials with two repulsive length scales have been identified as good
candidates for displaying novel glassy dynamics.  These models have been used to describe systems ranging from metallic glasses~\cite{Young77} to granular materials~\cite{duran}, as well as silica~\cite{FominSilica}, water~\cite{Jagla,Oliveira} and penetrable soft particles such as diblock copolymers, dendritic polymers, vesicles or microgels~\cite{Ziherl00,Pellicane03,Osterman07,ZiherlNat2014}.  
The square-shoulder (SS) model, i.e. a model system whose interaction potential is  a hard-core repulsion of extent $\sigma$ complemented by  a  repulsive shoulder, belongs to the family of core-softened potentials. For the SS model MCT calculations have predicted the existence of multiple glass transitions~\cite{Sperl2010}. Indeed, for specific values of the shoulder width $\Delta$, the  temperature $T$- packing fraction $\phi$ state  diagram is characterized by a non-monotonic fluid-glass transition line, retracing both upon cooling and upon compression.%~\cite{Sperl2010}.   
%Each reentrance separates two glasses with distinct properties. A first reentrance occurs at large packing fractions $\phi$ where two glasses, at high and low temperature $T$, are separated by a pocket of liquid states which are stabilized by the competition between the two localization lengths in the interaction potential. A second reentrance has been predicted at low $T$ and $\phi$.  In that case, the system can be considered as a renormalized hard-sphere (HS) glass with size $\sigma+\Delta$, since particles have not enough energy to overcome their mutual shoulder. The resulting glass can be melted upon compression and a second different glass can be obtained  at high $\phi$.
The peculiar MCT predictions for the SS is the additional presence of a  glass-glass line for small enough $\Delta$ terminating with two $A_3$ singularities,  fully embedded within the glass phase as shown in the schematic  Fig.~\ref{fig:MCTlines} (a).   
%Beyond the $A_3$  end-points,  the two glasses are indistinguishable, .  
Upon increasing $\Delta$, the glass-glass line progressively moves towards the fluid-glass one and the two eventually merge (Fig.\ref{fig:MCTlines} (b),(c)). When the $A_3$ point collides with the fluid-glass line, 
MCT predicts the presence of a higher-order singularity, named $A_4$. The dynamics close to higher order singularities differs from the standard fluid-glass scenario.  Instead of the characteristic two-step dynamics,   the decay of  the density correlators 
%(both self and collective)   
%$\Phi_q(t)=\langle \rho_{-q}(0)\rho_{q}(t)\rangle$
 shows   a logarithmic dependence on time $t$.
 %, e.g. $\Phi_q(t) \sim f_q-C_q \ln(t)$.
Correspondingly, the mean-squared displacement (MSD) 
%$ \langle\delta r^2\rangle$  % with $\delta r=r(t)-r(0)$
shows a subdiffusive  behaviour, i.e. $\sim t^{a}$ with $a<1$\cite{Gotzesperl}. 
Similar features have been observed also in theoretical studies on facilitated models\cite{SellittoJCP2013}, confirming the robustness of MCT results.

Despite the novelty of these MCT predictions, the observation of the peculiar dynamic behavior of the SS model, associated to higher-order singularities, has been difficult to achieve~\cite{Gayatri}. Indeed, the $A_3$ points are buried in the glass region and hence can not be accessed via an ergodic path.  Differently, the $A_4$ can  be approached from the fluid (ergodic) side and hence its role can be explored in equilibrium.  In addition, 
since the SS model is characterized by two $A_3$ points, it should have two distinct  $A_4$ points (see Fig.\ref{fig:MCTlines} (b),(c)).   
In this Letter we provide numerical evidence of the existence of two $A_4$ singularities in the SS model, by observing  subdiffusive behaviour of the MSD and logarithmic decay of the density correlators for several time decades. Additionally, we find a novel dynamic behaviour that occurs in the fluid region, generated by the interplay of the
two closeby higher-order points. We discover that their simultaneous presence
 gives rise to special loci in the $T-\phi$ plane along which the dynamics is identical ({\it iso-dynamics} loci), i.e. where both the short- and the long-time dynamics of the system remarkably coincide \textit{at all} length scales. 

{\it Model and Methods:}
We perform event-driven molecular dynamics simulations
of a $50:50$ non-crystallising binary mixture of $N=2000$ particles of species $A$ and $B$  interacting via the pairwise SS potential
\begin{equation}
V_{ij}(r)=\begin{cases}\infty, & r<\sigma_{ij}\\
u_0, & \sigma_{ij}\leq r<(1+\Delta) \sigma_{ij}\\
0, & r\geq (1+\Delta)\sigma_{ij},\\ 
\end{cases}
\end{equation}
\noindent where $i,j = A,B$, $\sigma_{ij}$ is the hard core between two particles, $\Delta$ is the shoulder width, and $u_0$ is the shoulder height.
The size ratio between the two species is $\sigma_{AA}/\sigma_{BB}=1.2$ and $\sigma_{AB}=(\sigma_{AA}+\sigma_{BB})/2$. The mass of particles is $m=1$.
% This slight difference in the species size allows to explore highly dense liquid state points with very low diffusivity without encountering crystallization, aiming to get as close as possible to the ideal liquid-glass line defined as the locus of state points  having diffusivity $D\rightarrow 0$.  
$\sigma_{BB}$ and $u_0$ are chosen as units of length and energy, so that the time $t$ is measured in units of
$t_0=\sigma_{BB} \sqrt{m/u_0}$.  $T$ is measured in units of energy ($k_B$=1). Simulations are performed in the canonical and microcanonical ensemble for a wide range of $T$ and $\phi=(\pi/6)(\rho_A\sigma^3_{AA}+\rho_{B}\sigma^3_{BB})$, where $\rho_A=
\rho_B=\rho/2$, $\rho \equiv N/V$ with $V$ the volume 
of the cubic simulation box. 

%We also solve the MCT for the monodisperse SS system finding the presence of two $A_4$ singularities within the Rogers-Young (RY) closure. These predictions have been confirmed by repeating the MCT calculations for our binary mixture, using as input of the theory the structure factors $S(q)$ evaluated in simulations, thus avoiding to rely on a specific closure of the Ornstein-Zernike equation. From these calculations we estimate $\Delta^{*(1)}_{MCT}\simeq 0.17$ and $\Delta^{*(2)}_{MCT}\simeq 0.20$ for the situations illustrated respectively in Fig.~\ref{fig:MCTlines}(b) and (c).

%The MCT dynamic phase diagram for $\Delta=\Delta^*$ is shown in Fig.\ref{fig:simulation-mct}.
 \begin{figure*}[t!]
\centering
\subfloat[]{\includegraphics[width=0.33\textwidth]{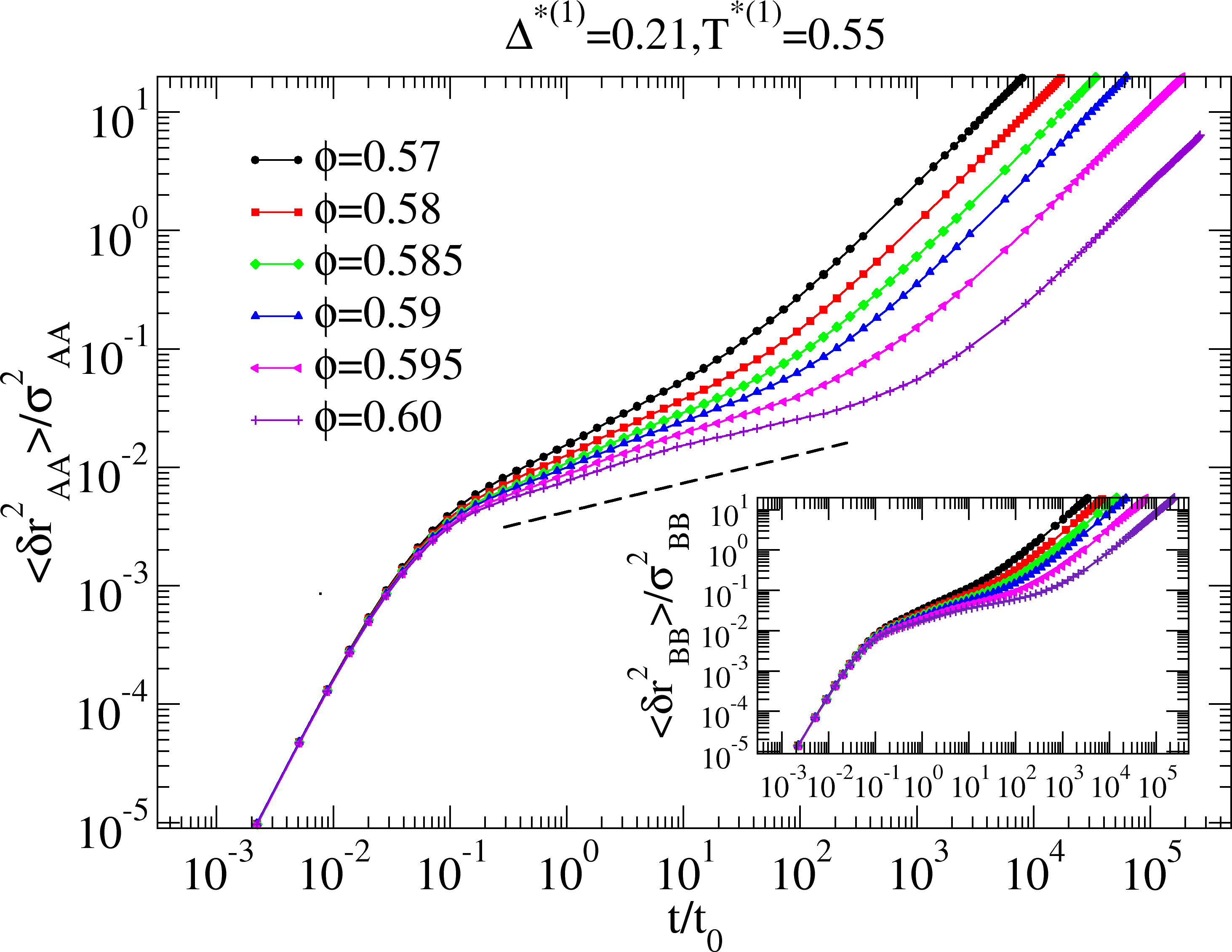}}
\subfloat[]{\includegraphics[width=0.33\textwidth]{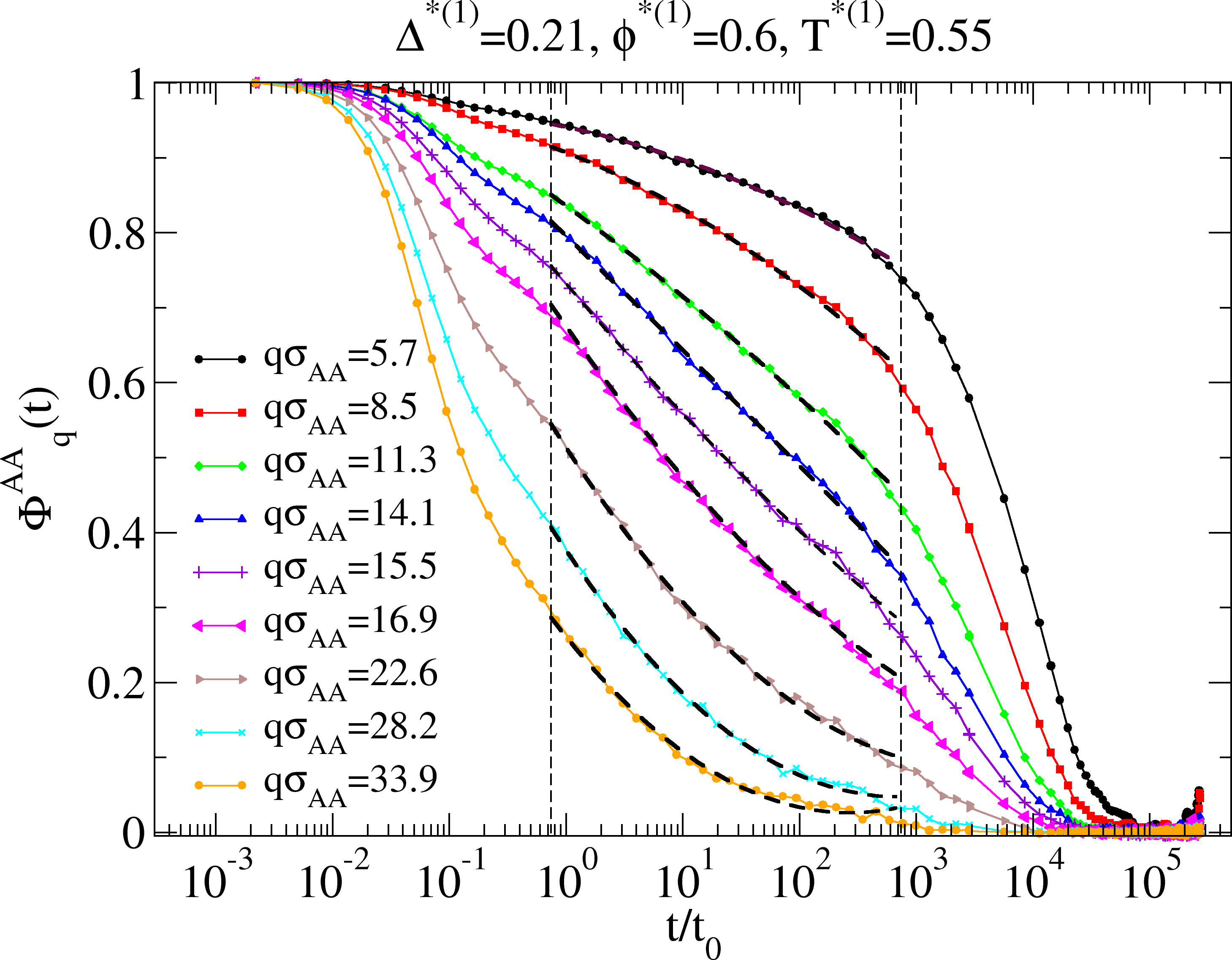}}\hspace{0.01cm}
\subfloat[]{\includegraphics[width=0.33\textwidth]{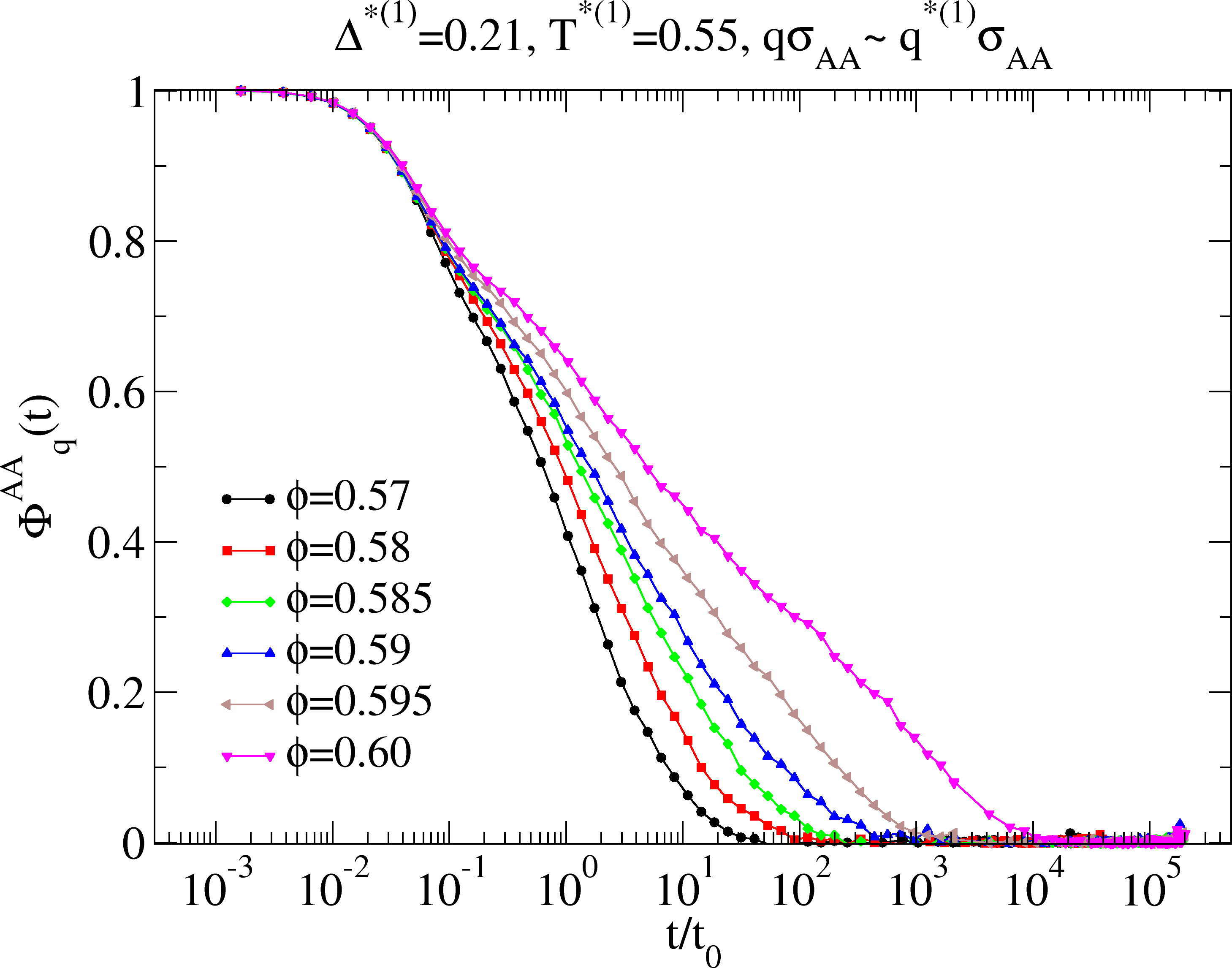}}\hspace{0.1cm}

\subfloat[]{\includegraphics[width=0.33\textwidth]{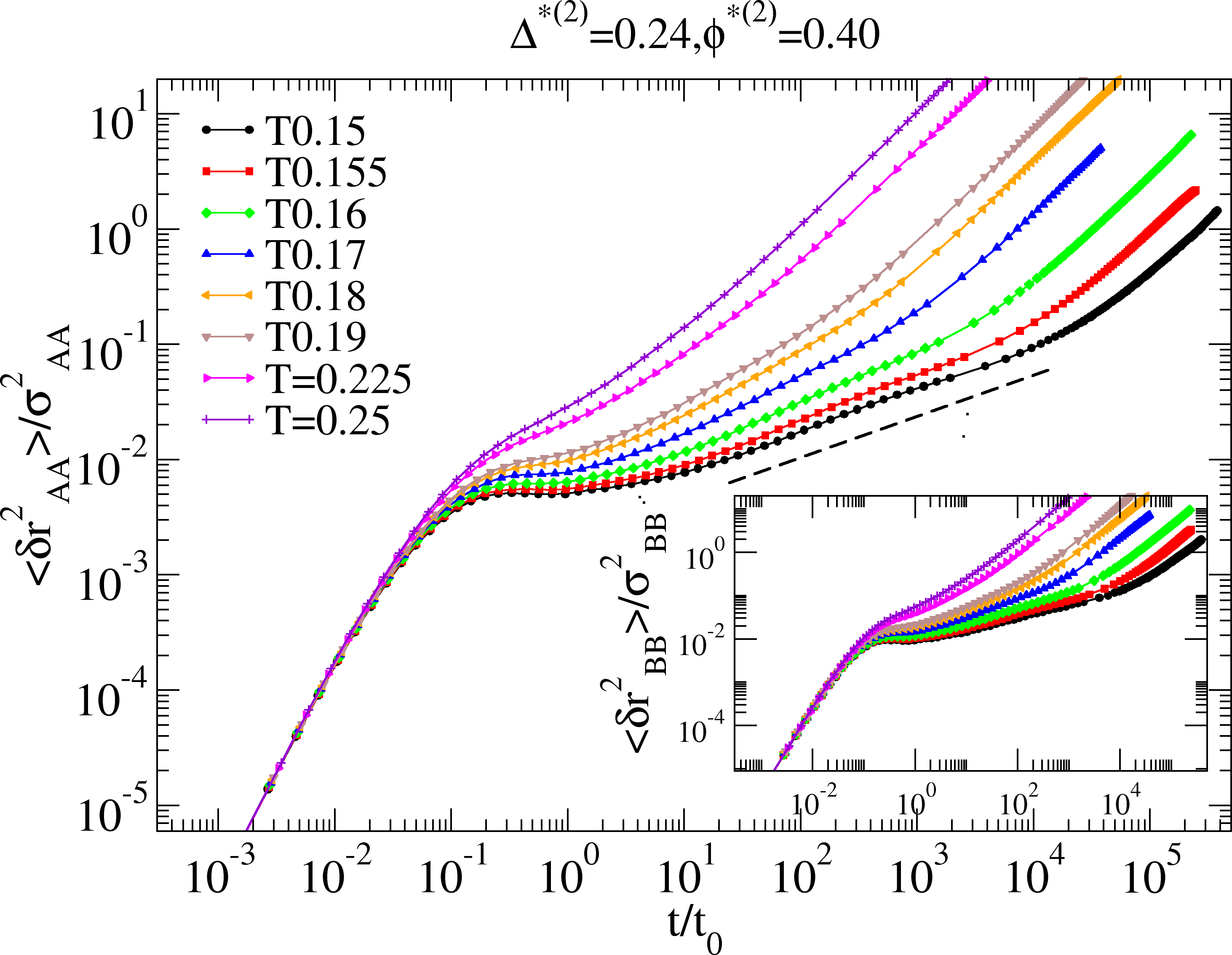}}
\subfloat[]{\includegraphics[width=0.33\textwidth]{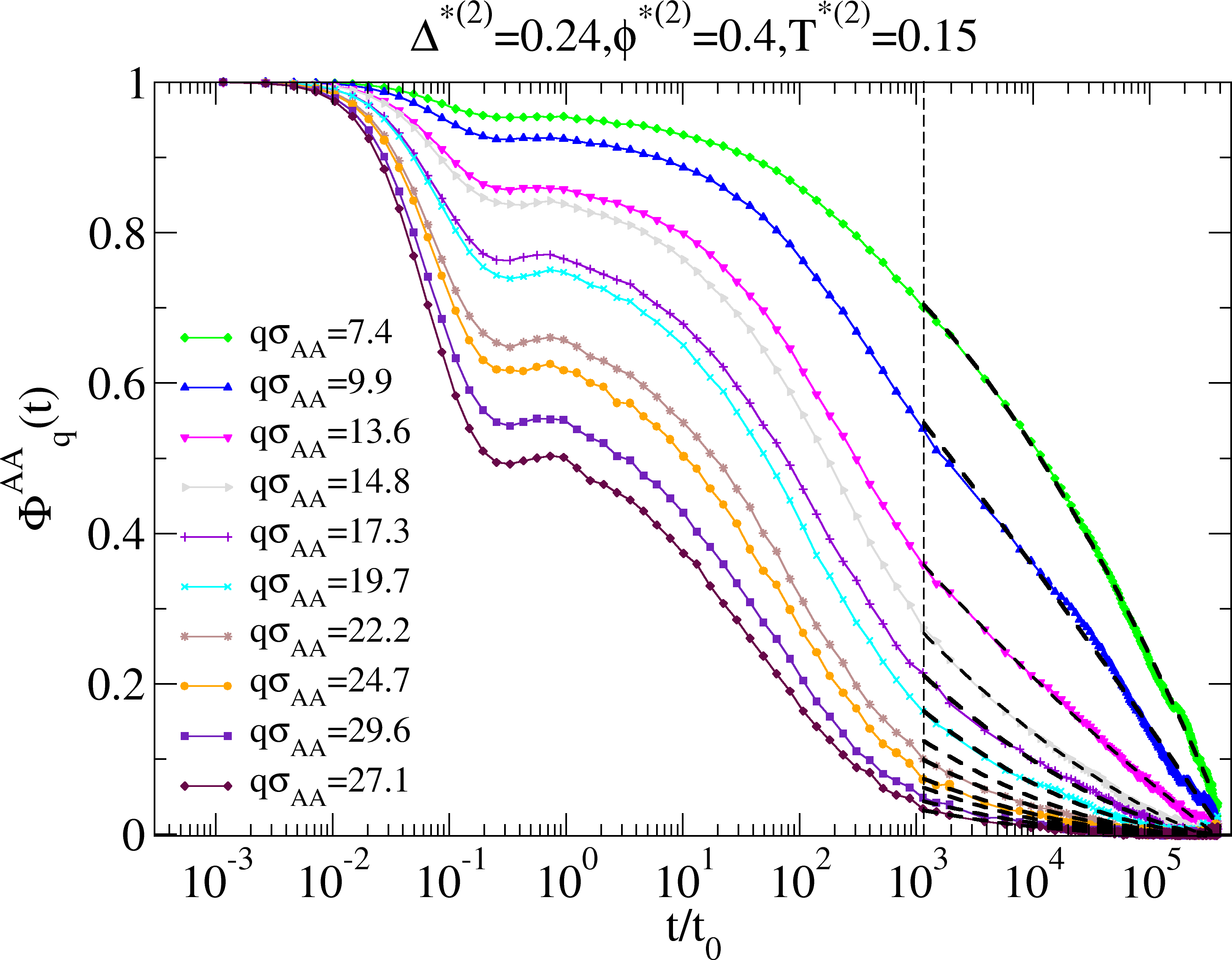}}\hspace{0.01cm}
\subfloat[]{\includegraphics[width=0.33\textwidth]{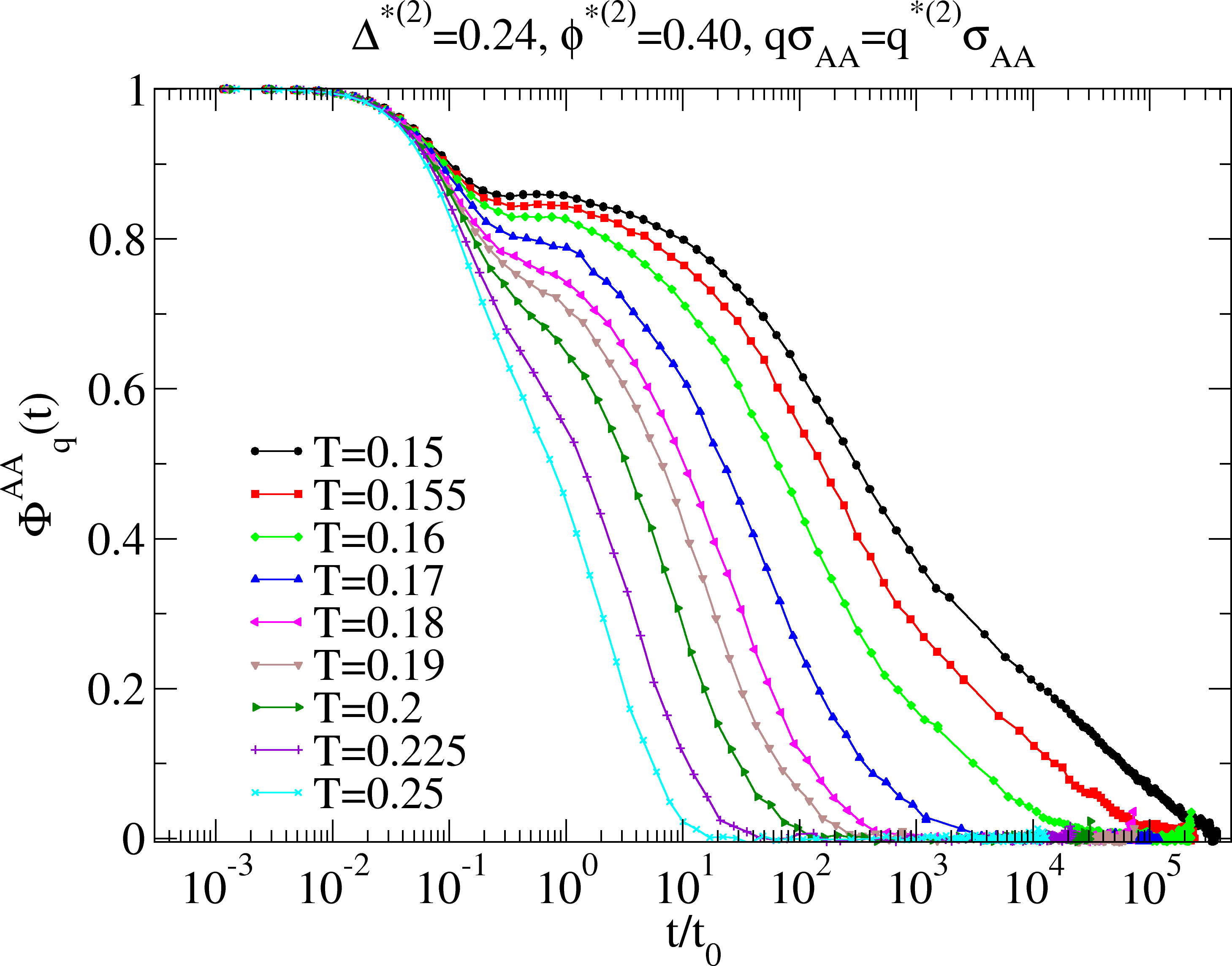}}
\caption{ Dynamical quantities  close to the two higher-order singularities $A_4^{(1)}\equiv(\phi^{*(1)}\simeq 0.6, T^{*(1)}\simeq 0.55,\Delta^{*(1)}\simeq 0.21)$ and $A_4^{(2)}\equiv (\phi^{*(2)}\simeq 0.4, T^{*(2)}\simeq 0.15,\Delta^{*(2)}\simeq 0.24)$, along paths shown in Fig.\protect\ref{fig:MCTlines} (b) and (c) respectively. (a) A-particles MSD  as a function of the scaled time $t_{0}$ for different $\phi$ at $T^{*(1)}, \Delta^{*(1)}$. The dashed line is a power-law  ($\propto t^{0.3}$)  to highlight the subdiffusive regime. The inset shows the MSD for the $B$ species.  (b) Collective density correlators of the A species $\Phi_q^{AA}(t)$ evaluated at  $A_4^{(1)}$ for different wave vectors $q\sigma_{AA}$. The dotted vertical lines delimit the time window in which $\Phi_q^{AA}(t)$  display a logarithmic behaviour. (c) $\Phi_{q}^{AA}(t)$ at $T^{*(1)}, \Delta^{*(1)}$ as a function of $\phi$. (d) MSD for A particles at $T^{*(2)}, \Delta^{*(2)}$  as function of the scaled time $t_{0}$ for different $T$. The subdiffusive regime  is characterised by $\propto t^{0.37}$. Inset: MSD of B particles. (e) The same as (b) but evaluated at the $A_4^{(2)}$.  Despite the interference of the fluid-glass line on the dynamics (see text), a long time  logarithmic behaviour can be identified. (f)$\Phi_{q^{*(2)}}^{AA}(t)$ at $\phi^{*(2)}, \Delta^{*(2)}$ as a function of $T$.}
\label{fig:A4}
\end{figure*}

%\noindent  
To locate the higher-order singularities we perform an extensive study of the dynamics of the 
SS system. Building on the previous study
for $\Delta=0.15$~\cite{Gayatri}, where comparison between simulations and theoretical predictions 
have provided an estimate of the $A_3$ points,  we restricted our search of the two $A_4$ points to values
of $\Delta \geq 0.17$. Specifically, we have analyzed in depth the range $0.17 \leq \Delta \leq 0.24$, with a mesh of $0.01$. For each $\Delta$ we have studied the dynamics in a wide window of $\phi$ and $T$.  Such  lengthy investigation  has allowed us to locate the first $A_4^{(1)}$ singularity at the state point ($\phi^{*(1)}\simeq 0.6$, $T^{*(1)}\simeq 0.55$, $\Delta^{*(1)}\simeq 0.21$) and  the second $A_4^{(2)}$ at  ($\phi^{*(2)}\simeq 0.4$, $T^{*(2)}\simeq 0.15$, $\Delta^{*(2)}\simeq 0.24$).  Details of this procedure, based on the  mapping of the MCT prediction onto the numerical data, are provided in the Supplementary Material. 

{\it Dynamics close to the two $A_4$ singularities:}
Figure \ref{fig:A4} shows the dynamic behaviour of the system close to the two  $A_4$ singularities. In the case of 
$A_4^{(1)}$ (top row panels) we follow the evolution of dynamic quantities on changing $\phi$ at $\Delta^{*(1)}$  and $T^{*(1)}$ fixed, while for $A_4^{(2)}$ (bottom row panels) we work at fixed $\Delta^{*(2)}$  and $\phi^{*(2)}$
upon  changing $T$, following the paths highlighted in Fig.~\ref{fig:MCTlines} (b) and (c).
%The reason for two different equilibrium paths for the two  points are clear by looking at the shapeof the glass lines reported in Fig.~\ref{fig:MCTlines}.   
Fig. \ref{fig:A4}(a) shows  the increasing subdiffusive behaviour of the MSD for  the A particles $\langle \delta r_{AA}^2\rangle$ on increasing $\phi$ which extends up to three orders of magnitude at $\phi=0.6$.  The B particles behave likewise (inset of Fig. \ref{fig:A4}(a)). The collective density correlators $\Phi^{AA}_{q}(t)$ of the $A$ particles display similarly striking features.
This is shown in Figure \ref{fig:A4} (b) where the correlators are reported for several wave-vectors $q\sigma_{AA}$. For a large time window 
the decay of  $\Phi^{AA}_{q}(t)$ is well described by a  second degree polynomial in $\ln(t)$, i.e.
$\Phi^{AA}_q(t) \sim f_q-H^{(1)}_q\ln(t/\tau)+H^{(2)}_q \ln(t/\tau)^2$, where $f_q$, $H^{(1)}_q$,$H^{(2)}_q$ are fit parameters. For a special wavevector $q^{*(1)}$, $H^{(2)}_q\sim 0$ and the correlator displays a pure logarithmic decay. We find $q^{*(1)}\sigma_{AA}=15.5$.
%: a clear $\ln(t)$ decay over a large time window is found.   
In addition, in the $q$-vector region explored we observe the  convex-to-concave crossover predicted by MCT\cite{Gotzesperl}. The $\phi$ dependence of  $\Phi^{AA}_{q}(t)$, reported in Fig. \ref{fig:A4} (c),  shows the growth of the logarithmic regime on approaching  $A_4^{(1)}$.

The same analysis has been carried out for $A_4^{(2)}$. As shown in Fig.\ref{fig:MCTlines} (c), such singularity lays close to the reentrance, which makes it difficult to explore the region around it by moving along constant $T$ paths. In addition, even along the constant $\phi$ path (i.e. by varying $T$)  the presence of a 
 re-entrant  fluid-glass line  (imposing its two-step relaxation behavior)  partially interferes with
 the logarithmic dynamics induced by the $A_4$ point.  As a result, dynamical quantities in Fig. \ref{fig:A4} (d)-(f) look  different from those obtained for the $A_4^{(1)}$.
Figure~\ref{fig:A4} (d) shows $\langle \delta r_{AA}^2\rangle$ as a function of $T$ for the $A$ and $B$ (inset) particles.   In this case the subdiffusive region is preceded by a plateau, a signature of the standard caging effect
imposed by the nearby  liquid-glass line. The subdiffusive region, the hallmark  of the higher-order singularity, is shifted to higher times  and extends over almost three decades.  The interplay between the fluid-glass line and
the $A_4$ dynamics is also observable in the decay of  $\Phi^{AA}_{q}(t)$  at ($\phi^{*(2)}$, $T^{*(2)}$, $\Delta^{*(2)}$) in Fig.~\ref{fig:A4}(e). For all  $q$-vectors the initial part of the structural relaxation displays the typical two-step behavior, but, for longer times, the decay becomes logarithmic.  We find a pure logarithmic behaviour in $\Phi^{AA}_{q}(t)$ for $q^{*(2)}\sigma_{AA}=13.6$. The  evolution  of  $\Phi^{AA}_{q}(t)$ at $q^{*(2)}\sigma_{AA}$ with $T$ is shown in Fig.~\ref{fig:A4}(f): the long time decay becomes more and more linear in $\ln(t)$ on approaching  $A_4^{(2)}$.

\begin{figure}
\centering
\includegraphics[width=0.4\textwidth]{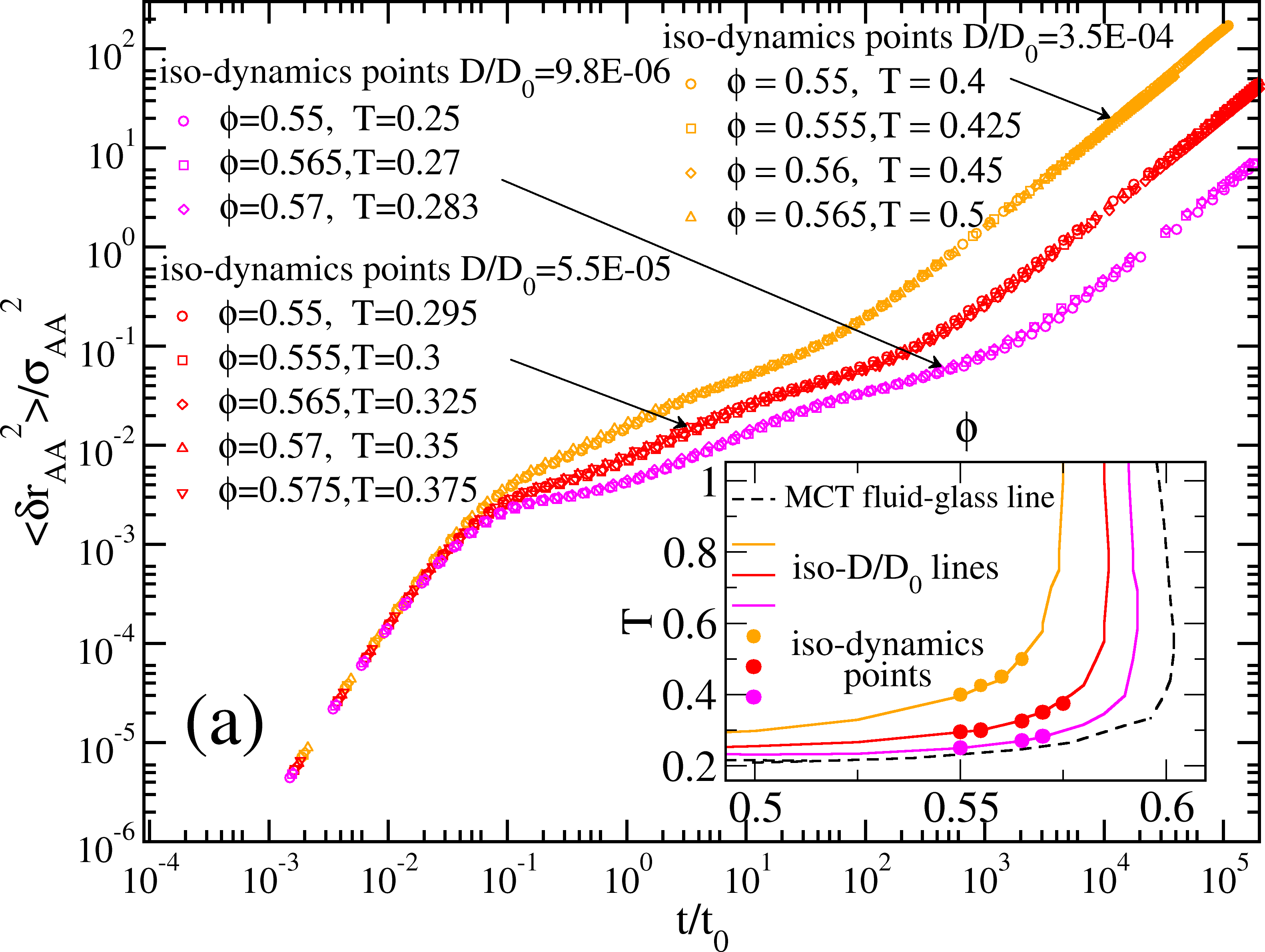}
\includegraphics[width=0.4\textwidth]{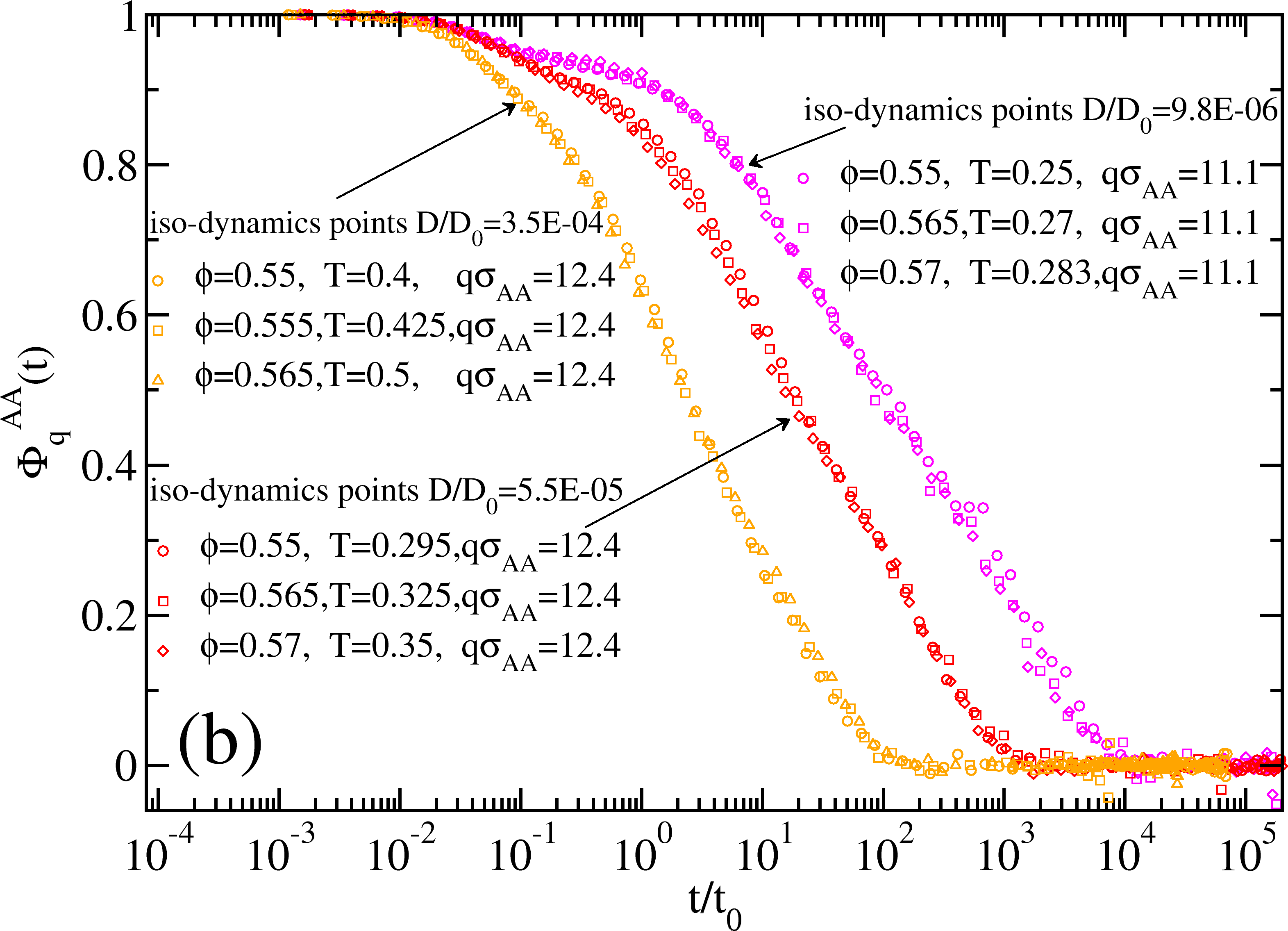}
\caption{Dynamical properties for $\Delta=0.17$ along three iso-dynamics lines with rescaled diffusivity $D/D_{0}=3.5\times10^{-4}$ (orange symbols), $D/D_{0}=5.5\times10^{-5}$ (red symbols) and $D/D_{0}=9.8\times10^{-6}$ (magenta symbols). (a) MSD of the $A$ species along the iso-dynamics line as a function of $t/t_{0}$ (b) Density correlator for the same sets of iso-dynamics points in (a) as a function of $t/t_{0}$. The inset shows the position of the iso-$D/D_{0}$ lines and of the expected fluid-glass line (see Supplementary Material).
%Closed symbols indicate iso-dynamics state points. The fluid-glass line has been calculated by solving the binary mixture MCT equations using as input the $S(q)$ from the simulation. A bilinear transformation in $T$ and $\rho$ has been applied to match the numerically determined  location of the glass line ().
}
\label{fig:Invariants}
\end{figure}

\begin{figure}
\centering
\includegraphics[width=0.4\textwidth]{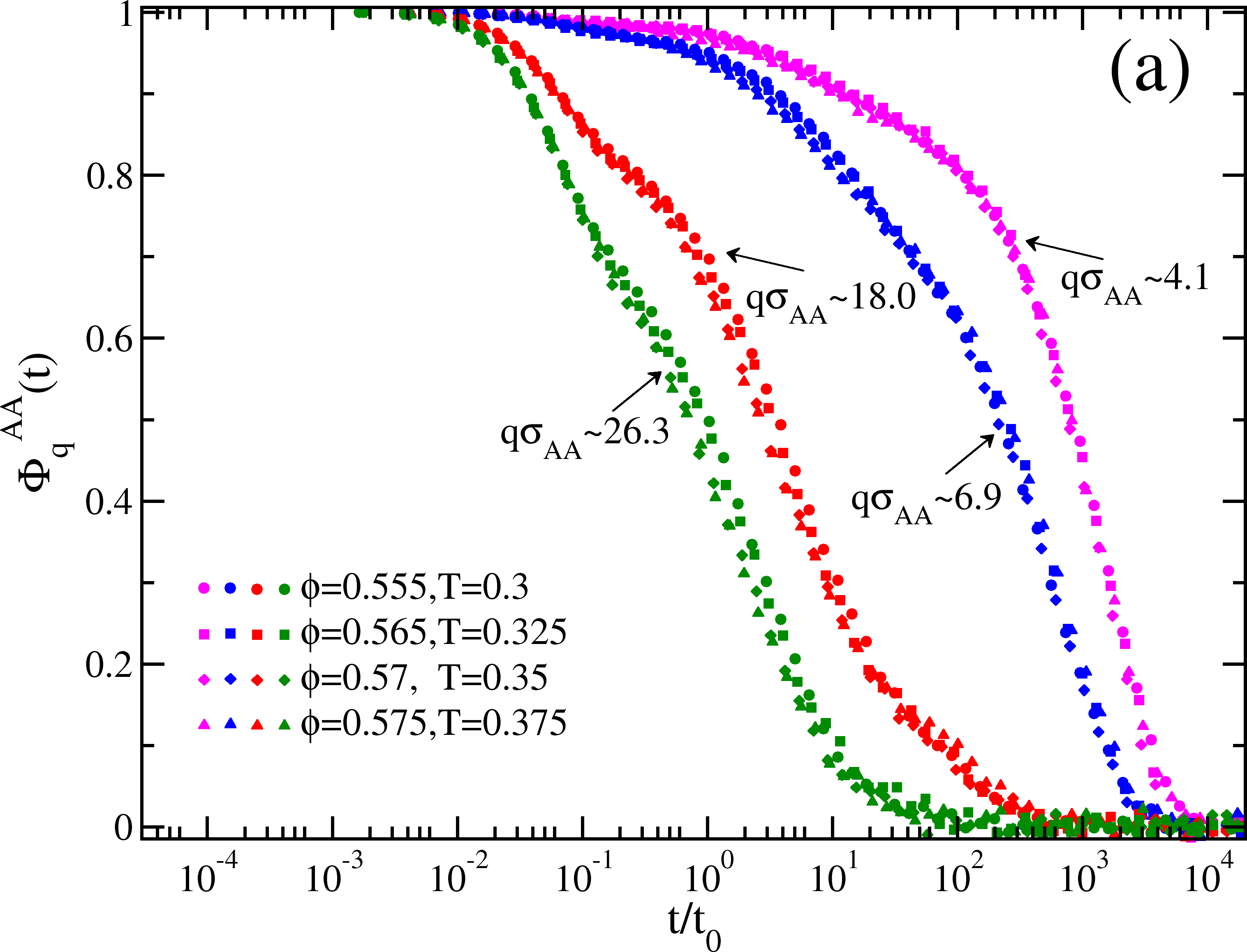}
\includegraphics[width=0.4\textwidth]{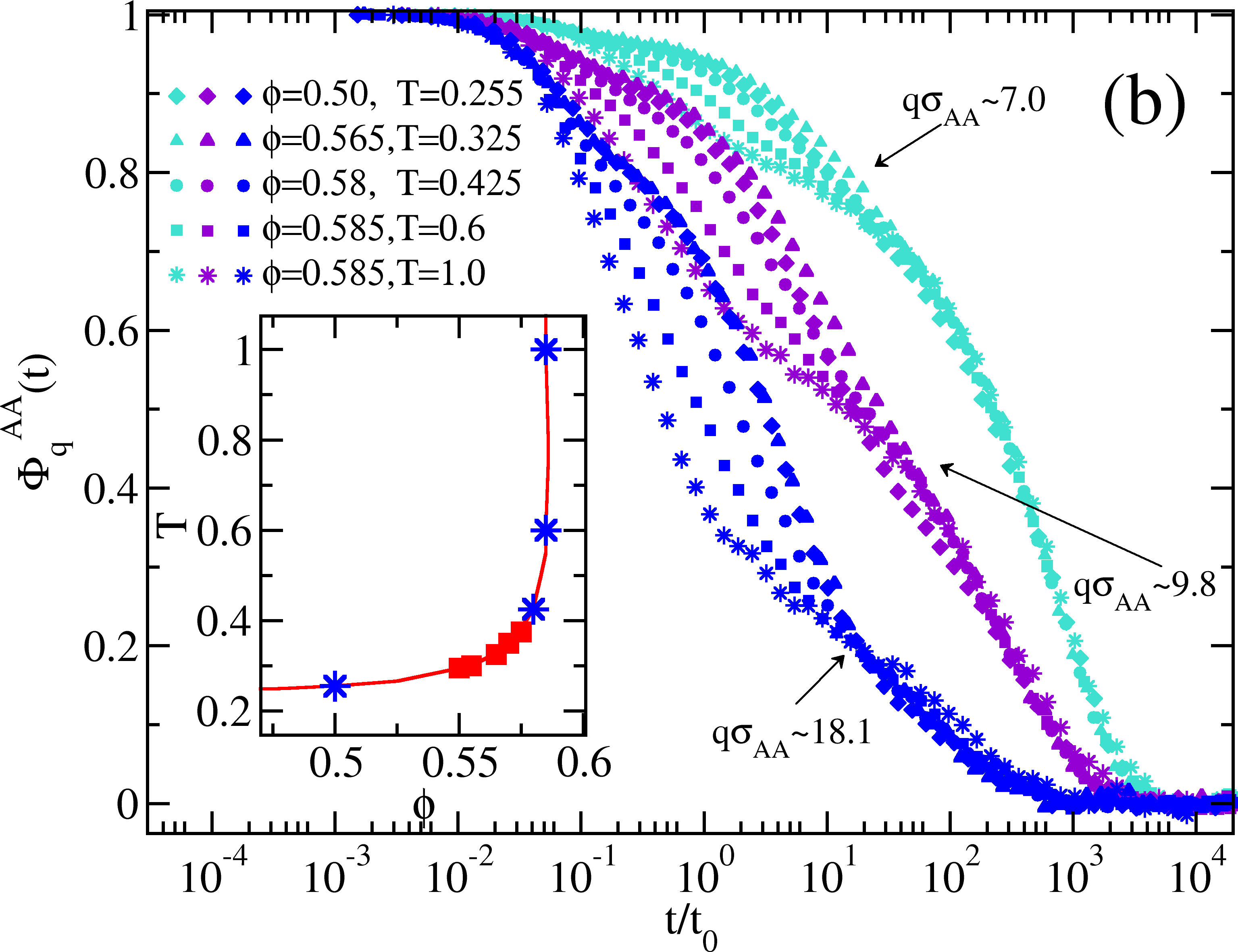}
\caption{ $\Phi^{AA}_{q}(t)$ along the iso-$D/D_{0}=5.5\times10^{-5}$ line for $\Delta=0.17$ at different wave vectors $q\sigma_{AA}$ for (a) iso-dynamics and (b) non iso-dynamics state points. The inset shows the
location of such points in the $T$-$\phi$ diagram. Squares/stars  indicate iso-dynamics/non iso-dynamics points.
}
\label{fig:FQTInvariants}
\end{figure}

%Although MCT is able to describe how the dynamical quantities change in the presence of a higher-order singularity, it does not give any hints on how the dynamics of a system could be affected when multiple higher-order singularities influence a given region of the state diagram. This is the case of the SS system, in which ,depending on the value of $\Delta$, $A_3$ or $A_4$ points influence simultaneously the dynamics of states laying in between the two singularities. In such a case, the competition between the two could generate a peculiar and non-trivial dynamics that has never been observed so far.

{\it Iso-dynamics lines:}  The procedure to locate the $A_4$ singularities required the investigation of a very large number of state points for several $\Delta$ values.  During such process, we have discovered a peculiar dynamic feature of the SS
model that we associate to the simultaneous presence of two distinct higher order singularities ($A_3$ or $A_4$).
Specifically, we find that  the competition between these two special points generates loci in the $T$-$\phi$ plane with invariant dynamics, that we name iso-dynamics lines. To gain a deeper understanding of such loci we investigate in details 
the region  in between two $A_3$  singularities for $\Delta=0.17$, i.e. for the case  schematically shown in
Fig.~\ref{fig:MCTlines}(b). % (for different $\Delta$ values  see Supplementary Material). 
We focus on iso-diffusivity (iso-$D/D_0$)\cite{Zac02a} paths close to the fluid-glass line  ($D_0 \equiv \sigma_{BB}^2/t_0$
accounts for the trivial effect of the thermal velocity), as shown in the inset of Fig.\ref{fig:Invariants}(a).
%To detect the presence of a special dynamics in the region between the two singularities we have investigated the dynamical properties of a number of state points lying on different iso-diffusivity (iso-$D/D_0$) lines for $\Delta=0.17$ with intermediate and low $D/D_0$ as shown in Fig. \ref{fig:iso-invariant}. 
%As discussed above, for $\Delta=0.17$ there are two $A_3$ singularities so close to the liquid-glass line to clearly observe their influence on a wide region of points in the dynamical state diagram. Surprisingly, 
Surprisingly, we find that state points along iso-$D/D_0$ curves are characterized not only by the same long-time dynamics but also by the same 
short and intermediate time dependence. 
Figures \ref{fig:Invariants}(a) and (b) show, respectively, the MSD and $\Phi^{AA}_{q}(t)$ as a function of  $t/t_0$, along  three iso-$D/D_0$ lines, differing by more than two orders of magnitude in $D/D_0$. 
%with $D/D_{0}=3.5\times10^{-4}$, $D/D_{0}=5.5\times10^{-5}$ and $D/D_{0}=9.8\times10^{-6}$.  
For all three  iso-$D/D_0$ sets, the superposition of the curves, {\it at all times}  and {\it at all length scales} is striking,
both in real and in Fourier space.   To further support the iso-dynamics behavior at all length scales
we show in Fig.~\ref{fig:FQTInvariants}(a) the wave-vector dependence of $\Phi^{AA}_{q}(t)$ for a specific value
of $D/D_0$.     Again,  superposition of the correlation functions at all times is observed for all $q$ values.
These results prove the existence of  iso-dynamics loci, i.e. lines in the $T-\phi$ plane where an identical dynamics is observed. It is interesting to notice that while dynamics is identical, structural and thermodynamic properties are not, as discussed in the Supplementary Material. 
Finally we remark that only   state points  that feel the presence of both higher order singularities  obey the invariance.  Fig.~\ref{fig:FQTInvariants}(b)  shows that  the decay of $\Phi^{AA}_{q}(t)$ does not satisfy the invariance for $T$ and $\phi$ progressively moving closer to one  singularity  (but always on the iso-$D/D_0$ line).  
In this case, while the long time dynamics  is identical (as one could  expect on the basis of the identical diffusion coefficient), the short and intermediate time dynamics is now clearly different, indicating that the system explores the nearest-neighbour cages in a different way for each state point.   The state points where the iso-dynamics is/is-not observed are indicated in the inset of Fig.~\ref{fig:FQTInvariants}(b). In the Supplementary Material, we discuss how the iso-dynamics behaviour is compatible with MCT predictions.

 \emph{Conclusions:} We have reported  numerical evidence of the existence of two $A_4$ singularities in a simple system with two repulsive length scales. We have confirmed that anomalous dynamical features, such as the logarithmic decay of the density autocorrelation function and the subdiffusive regime in the mean-square displacement, characterize the dynamics close to these points. This result provides (i) one of the most stringent test of previously formulated  MCT predictions (ii) evidence that soft-colloidal particles could  constitute a model system for experimentally testing these highly unconventional behavior.  In addition, we unexpectedly discovered that the competition between two higher-order singularities  gives rise to a non-trivial  iso-dynamics in between the two singularities. Such state points share the same dynamics at all time and length-scales.  We hope our study will stimulate the experimental search for anomalous dynamics and competing glass transitions in core-softened systems and in repulsive systems with two competing length scales.

\emph{Acknowledgments} NG, EZ acknowledge support from MIUR ("Futuro in Ricerca" ANISOFT/RBFR125H0M); GD, FS, EZ from EU ITN-234810-COMPLOIDS.

%\bibliography{./articoli}

\begin{thebibliography}{36}
\expandafter\ifx\csname natexlab\endcsname\relax\def\natexlab#1{#1}\fi
\expandafter\ifx\csname bibnamefont\endcsname\relax
  \def\bibnamefont#1{#1}\fi
\expandafter\ifx\csname bibfnamefont\endcsname\relax
  \def\bibfnamefont#1{#1}\fi
\expandafter\ifx\csname citenamefont\endcsname\relax
  \def\citenamefont#1{#1}\fi
\expandafter\ifx\csname url\endcsname\relax
  \def\url#1{\texttt{#1}}\fi
\expandafter\ifx\csname urlprefix\endcsname\relax\def\urlprefix{URL }\fi
\providecommand{\bibinfo}[2]{#2}
\providecommand{\eprint}[2][]{\url{#2}}

\bibitem[{\citenamefont{Mallamace et~al.}(2000)\citenamefont{Mallamace,
  Gambadauro, Micali, Tartaglia, Liao, and Chen}}]{Mal00a}
\bibinfo{author}{\bibfnamefont{F.}~\bibnamefont{Mallamace}},
  \bibinfo{author}{\bibfnamefont{P.}~\bibnamefont{Gambadauro}},
  \bibinfo{author}{\bibfnamefont{N.}~\bibnamefont{Micali}},
  \bibinfo{author}{\bibfnamefont{P.}~\bibnamefont{Tartaglia}},
  \bibinfo{author}{\bibfnamefont{C.}~\bibnamefont{Liao}}, \bibnamefont{and}
  \bibinfo{author}{\bibfnamefont{S.~H.} \bibnamefont{Chen}},
  \bibinfo{journal}{Phys. Rev. Lett.} \textbf{\bibinfo{volume}{84}},
  \bibinfo{pages}{5431} (\bibinfo{year}{2000}).

\bibitem[{\citenamefont{Eckert and Bartsch}(2002)}]{Eck02a}
\bibinfo{author}{\bibfnamefont{T.}~\bibnamefont{Eckert}} \bibnamefont{and}
  \bibinfo{author}{\bibfnamefont{E.}~\bibnamefont{Bartsch}},
  \bibinfo{journal}{Phys. Rev. Lett.} \textbf{\bibinfo{volume}{89}},
  \bibinfo{pages}{125701} (\bibinfo{year}{2002}).

\bibitem[{\citenamefont{Pham et~al.}(2002)\citenamefont{Pham, Puertas,
  Bergenholtz, Egelhaaf, Moussa{\" i}d, Pusey, Schofield, Cates, Fuchs, and
  Poon}}]{Pha02a}
\bibinfo{author}{\bibfnamefont{K.~N.} \bibnamefont{Pham}},
  \bibinfo{author}{\bibfnamefont{A.~M.} \bibnamefont{Puertas}},
  \bibinfo{author}{\bibfnamefont{J.}~\bibnamefont{Bergenholtz}},
  \bibinfo{author}{\bibfnamefont{S.~U.} \bibnamefont{Egelhaaf}},
  \bibinfo{author}{\bibfnamefont{A.}~\bibnamefont{Moussa{\" i}d}},
  \bibinfo{author}{\bibfnamefont{P.~N.} \bibnamefont{Pusey}},
  \bibinfo{author}{\bibfnamefont{A.~B.} \bibnamefont{Schofield}},
  \bibinfo{author}{\bibfnamefont{M.~E.} \bibnamefont{Cates}},
  \bibinfo{author}{\bibfnamefont{M.}~\bibnamefont{Fuchs}}, \bibnamefont{and}
  \bibinfo{author}{\bibfnamefont{W.~C.~K.} \bibnamefont{Poon}},
  \bibinfo{journal}{Science} \textbf{\bibinfo{volume}{296}},
  \bibinfo{pages}{104} (\bibinfo{year}{2002}).

\bibitem[{\citenamefont{Chen et~al.}(2003)\citenamefont{Chen, W.-R.Chen, and
  Mallamace}}]{Chen03a}
\bibinfo{author}{\bibfnamefont{S.~H.} \bibnamefont{Chen}},
  \bibinfo{author}{\bibnamefont{W.-R.Chen}}, \bibnamefont{and}
  \bibinfo{author}{\bibfnamefont{F.}~\bibnamefont{Mallamace}},
  \bibinfo{journal}{Science} \textbf{\bibinfo{volume}{300}},
  \bibinfo{pages}{619} (\bibinfo{year}{2003}).

\bibitem[{\citenamefont{Pham et~al.}(2004)\citenamefont{Pham, Egelhaaf, Pusey,
  and Poon}}]{Pha04a}
\bibinfo{author}{\bibfnamefont{K.~N.} \bibnamefont{Pham}},
  \bibinfo{author}{\bibfnamefont{S.~U.} \bibnamefont{Egelhaaf}},
  \bibinfo{author}{\bibfnamefont{P.~N.} \bibnamefont{Pusey}}, \bibnamefont{and}
  \bibinfo{author}{\bibfnamefont{W.~C.~K.} \bibnamefont{Poon}},
  \bibinfo{journal}{Phys. Rev. E} \textbf{\bibinfo{volume}{69}},
  \bibinfo{pages}{1} (\bibinfo{year}{2004}).

\bibitem[{\citenamefont{Lu et~al.}(2008)\citenamefont{Lu, Mochrie, Narayanan,
  Sandy, and Sprung}}]{Sprung08}
\bibinfo{author}{\bibfnamefont{X.}~\bibnamefont{Lu}},
  \bibinfo{author}{\bibfnamefont{S.~G.~J.} \bibnamefont{Mochrie}},
  \bibinfo{author}{\bibfnamefont{S.}~\bibnamefont{Narayanan}},
  \bibinfo{author}{\bibfnamefont{A.~R.} \bibnamefont{Sandy}}, \bibnamefont{and}
  \bibinfo{author}{\bibfnamefont{M.}~\bibnamefont{Sprung}},
  \bibinfo{journal}{Phys. Rev. Lett.} \textbf{\bibinfo{volume}{100}},
  \bibinfo{pages}{045701} (\bibinfo{year}{2008}).

\bibitem[{\citenamefont{G\"{o}tze}(2009)}]{Gotzebook}
\bibinfo{author}{\bibfnamefont{W.}~\bibnamefont{G\"{o}tze}},
  \emph{\bibinfo{title}{Complex Dynamics of Glass-Forming Liquids: A
  Mode-Coupling Theory}} (\bibinfo{publisher}{Oxford Univ Press, New York},
  \bibinfo{year}{2009}).

\bibitem[{\citenamefont{Fabbian et~al.}(1999)\citenamefont{Fabbian, G{\" o}tze,
  Sciortino, Tartaglia, and Thiery}}]{Fab99a}
\bibinfo{author}{\bibfnamefont{L.}~\bibnamefont{Fabbian}},
  \bibinfo{author}{\bibfnamefont{W.}~\bibnamefont{G{\" o}tze}},
  \bibinfo{author}{\bibfnamefont{F.}~\bibnamefont{Sciortino}},
  \bibinfo{author}{\bibfnamefont{P.}~\bibnamefont{Tartaglia}},
  \bibnamefont{and} \bibinfo{author}{\bibfnamefont{F.}~\bibnamefont{Thiery}},
  \bibinfo{journal}{Phys. Rev. E} \textbf{\bibinfo{volume}{59}},
  \bibinfo{pages}{1347} (\bibinfo{year}{1999}).

\bibitem[{\citenamefont{Bergenholtz and Fuchs}(1999)}]{Ber99a}
\bibinfo{author}{\bibfnamefont{J.}~\bibnamefont{Bergenholtz}} \bibnamefont{and}
  \bibinfo{author}{\bibfnamefont{M.}~\bibnamefont{Fuchs}},
  \bibinfo{journal}{Phys. Rev. E} \textbf{\bibinfo{volume}{59}},
  \bibinfo{pages}{5706} (\bibinfo{year}{1999}).

\bibitem[{\citenamefont{Dawson et~al.}(2001)\citenamefont{Dawson, Foffi, Fuchs,
  G\"otze, Sciortino, Sperl, Tartaglia, Voigtmann, and Zaccarelli}}]{Daw00a}
\bibinfo{author}{\bibfnamefont{K.~A.} \bibnamefont{Dawson}},
  \bibinfo{author}{\bibfnamefont{G.}~\bibnamefont{Foffi}},
  \bibinfo{author}{\bibfnamefont{M.}~\bibnamefont{Fuchs}},
  \bibinfo{author}{\bibfnamefont{W.}~\bibnamefont{G\"otze}},
  \bibinfo{author}{\bibfnamefont{F.}~\bibnamefont{Sciortino}},
  \bibinfo{author}{\bibfnamefont{M.}~\bibnamefont{Sperl}},
  \bibinfo{author}{\bibfnamefont{P.}~\bibnamefont{Tartaglia}},
  \bibinfo{author}{\bibfnamefont{T.}~\bibnamefont{Voigtmann}},
  \bibnamefont{and}
  \bibinfo{author}{\bibfnamefont{E.}~\bibnamefont{Zaccarelli}},
  \bibinfo{journal}{Phys. Rev. E} \textbf{\bibinfo{volume}{63}},
  \bibinfo{pages}{011401} (\bibinfo{year}{2001}).

\bibitem[{\citenamefont{G\"{o}tze and Sperl}(2002)}]{Gotzesperl}
\bibinfo{author}{\bibfnamefont{W.}~\bibnamefont{G\"{o}tze}} \bibnamefont{and}
  \bibinfo{author}{\bibfnamefont{M.}~\bibnamefont{Sperl}},
  \bibinfo{journal}{Phys. Rev. E} \textbf{\bibinfo{volume}{66}},
  \bibinfo{pages}{011405} (\bibinfo{year}{2002}).

\bibitem[{\citenamefont{Puertas et~al.}(2002)\citenamefont{Puertas, Fuchs, and
  Cates}}]{Pue02a}
\bibinfo{author}{\bibfnamefont{A.~M.} \bibnamefont{Puertas}},
  \bibinfo{author}{\bibfnamefont{M.}~\bibnamefont{Fuchs}}, \bibnamefont{and}
  \bibinfo{author}{\bibfnamefont{M.~E.} \bibnamefont{Cates}},
  \bibinfo{journal}{Phys. Rev. Lett.} \textbf{\bibinfo{volume}{88}},
  \bibinfo{pages}{098301} (\bibinfo{year}{2002}).

\bibitem[{\citenamefont{Foffi et~al.}(2002)\citenamefont{Foffi, Dawson,
  Buldyrev, Sciortino, Zaccarelli, and Tartaglia}}]{Fof02a}
\bibinfo{author}{\bibfnamefont{G.}~\bibnamefont{Foffi}},
  \bibinfo{author}{\bibfnamefont{K.~A.} \bibnamefont{Dawson}},
  \bibinfo{author}{\bibfnamefont{S.~V.} \bibnamefont{Buldyrev}},
  \bibinfo{author}{\bibfnamefont{F.}~\bibnamefont{Sciortino}},
  \bibinfo{author}{\bibfnamefont{E.}~\bibnamefont{Zaccarelli}},
  \bibnamefont{and}
  \bibinfo{author}{\bibfnamefont{P.}~\bibnamefont{Tartaglia}},
  \bibinfo{journal}{Phys. Rev. E} \textbf{\bibinfo{volume}{65}},
  \bibinfo{pages}{050802} (\bibinfo{year}{2002}).

\bibitem[{\citenamefont{Zaccarelli et~al.}(2002)\citenamefont{Zaccarelli,
  Foffi, Dawson, Buldyrev, Sciortino, and Tartaglia}}]{Zac02a}
\bibinfo{author}{\bibfnamefont{E.}~\bibnamefont{Zaccarelli}},
  \bibinfo{author}{\bibfnamefont{G.}~\bibnamefont{Foffi}},
  \bibinfo{author}{\bibfnamefont{K.~A.} \bibnamefont{Dawson}},
  \bibinfo{author}{\bibfnamefont{S.~V.} \bibnamefont{Buldyrev}},
  \bibinfo{author}{\bibfnamefont{F.}~\bibnamefont{Sciortino}},
  \bibnamefont{and}
  \bibinfo{author}{\bibfnamefont{P.}~\bibnamefont{Tartaglia}},
  \bibinfo{journal}{Phys. Rev. E} \textbf{\bibinfo{volume}{66}},
  \bibinfo{pages}{041402} (\bibinfo{year}{2002}).

\bibitem[{\citenamefont{Puertas et~al.}(2003)\citenamefont{Puertas, Fuchs, and
  Cates}}]{Pue03a}
\bibinfo{author}{\bibfnamefont{A.}~\bibnamefont{Puertas}},
  \bibinfo{author}{\bibfnamefont{M.}~\bibnamefont{Fuchs}}, \bibnamefont{and}
  \bibinfo{author}{\bibfnamefont{M.~E.} \bibnamefont{Cates}},
  \bibinfo{journal}{Phys. Rev. E} \textbf{\bibinfo{volume}{67}},
  \bibinfo{pages}{031406} (\bibinfo{year}{2003}).

\bibitem[{\citenamefont{Sciortino et~al.}(2003)\citenamefont{Sciortino,
  Tartaglia, and Zaccarelli}}]{Sci03a}
\bibinfo{author}{\bibfnamefont{F.}~\bibnamefont{Sciortino}},
  \bibinfo{author}{\bibfnamefont{P.}~\bibnamefont{Tartaglia}},
  \bibnamefont{and}
  \bibinfo{author}{\bibfnamefont{E.}~\bibnamefont{Zaccarelli}},
  \bibinfo{journal}{Phys. Rev. Lett.} \textbf{\bibinfo{volume}{91}},
  \bibinfo{pages}{268301} (\bibinfo{year}{2003}).

\bibitem[{\citenamefont{Zaccarelli et~al.}(2004)\citenamefont{Zaccarelli,
  Sciortino, and Tartaglia}}]{Zac04b}
\bibinfo{author}{\bibfnamefont{E.}~\bibnamefont{Zaccarelli}},
  \bibinfo{author}{\bibfnamefont{F.}~\bibnamefont{Sciortino}},
  \bibnamefont{and}
  \bibinfo{author}{\bibfnamefont{P.}~\bibnamefont{Tartaglia}},
  \bibinfo{journal}{J. Phys.: Condens. Matter} \textbf{\bibinfo{volume}{16}},
  \bibinfo{pages}{4849} (\bibinfo{year}{2004}).

\bibitem[{\citenamefont{Imhof and Dhont}(1995)}]{imhof95}
\bibinfo{author}{\bibfnamefont{A.}~\bibnamefont{Imhof}} \bibnamefont{and}
  \bibinfo{author}{\bibfnamefont{J.~K.~G.} \bibnamefont{Dhont}},
  \bibinfo{journal}{Phys. Rev. Lett.} \textbf{\bibinfo{volume}{75}},
  \bibinfo{pages}{1662} (\bibinfo{year}{1995}).

\bibitem[{\citenamefont{{Voigtmann}}(2011)}]{Voigt2011}
\bibinfo{author}{\bibfnamefont{T.}~\bibnamefont{{Voigtmann}}},
  \bibinfo{journal}{Europhys. Lett.} \textbf{\bibinfo{volume}{96}},
  \bibinfo{pages}{36006} (\bibinfo{year}{2011}).

\bibitem[{\citenamefont{{Moreno} and
  {Colmenero}}(2006{\natexlab{a}})}]{Moreno06PRE}
\bibinfo{author}{\bibfnamefont{A.~J.} \bibnamefont{{Moreno}}} \bibnamefont{and}
  \bibinfo{author}{\bibfnamefont{J.}~\bibnamefont{{Colmenero}}},
  \bibinfo{journal}{Phys. Rev. E} \textbf{\bibinfo{volume}{74}},
  \bibinfo{pages}{021409} (\bibinfo{year}{2006}{\natexlab{a}}).

\bibitem[{\citenamefont{{Moreno} and
  {Colmenero}}(2006{\natexlab{b}})}]{Moreno06JCP}
\bibinfo{author}{\bibfnamefont{A.~J.} \bibnamefont{{Moreno}}} \bibnamefont{and}
  \bibinfo{author}{\bibfnamefont{J.}~\bibnamefont{{Colmenero}}},
  \bibinfo{journal}{J. Chem. Phys.} \textbf{\bibinfo{volume}{125}},
  \bibinfo{pages}{164507} (\bibinfo{year}{2006}{\natexlab{b}}).

\bibitem[{\citenamefont{{Voigtmann} and {Horbach}}(2009)}]{voigtmannhorbach}
\bibinfo{author}{\bibfnamefont{T.}~\bibnamefont{{Voigtmann}}} \bibnamefont{and}
  \bibinfo{author}{\bibfnamefont{J.}~\bibnamefont{{Horbach}}},
  \bibinfo{journal}{Phys. Rev. Lett.} \textbf{\bibinfo{volume}{103}},
  \bibinfo{pages}{205901} (\bibinfo{year}{2009}).

\bibitem[{\citenamefont{Mayer et~al.}(2008)\citenamefont{Mayer, Zaccarelli,
  Stiakakis, Likos, Sciortino, Munam, Gauthier, Hadjichristidis, Iatrou,
  Tartaglia et~al.}}]{MayerNature}
\bibinfo{author}{\bibfnamefont{C.}~\bibnamefont{Mayer}},
  \bibinfo{author}{\bibfnamefont{E.}~\bibnamefont{Zaccarelli}},
  \bibinfo{author}{\bibfnamefont{E.}~\bibnamefont{Stiakakis}},
  \bibinfo{author}{\bibfnamefont{C.~N.} \bibnamefont{Likos}},
  \bibinfo{author}{\bibfnamefont{F.}~\bibnamefont{Sciortino}},
  \bibinfo{author}{\bibfnamefont{A.}~\bibnamefont{Munam}},
  \bibinfo{author}{\bibfnamefont{M.}~\bibnamefont{Gauthier}},
  \bibinfo{author}{\bibfnamefont{N.}~\bibnamefont{Hadjichristidis}},
  \bibinfo{author}{\bibfnamefont{H.}~\bibnamefont{Iatrou}},
  \bibinfo{author}{\bibfnamefont{P.}~\bibnamefont{Tartaglia}},
  \bibnamefont{et~al.}, \bibinfo{journal}{Nat. Mater.}
  \textbf{\bibinfo{volume}{7}}, \bibinfo{pages}{780} (\bibinfo{year}{2008}).

\bibitem[{\citenamefont{Mayer et~al.}(2009)\citenamefont{Mayer, Sciortino,
  Likos, Tartaglia, Loewen, and Zaccarelli}}]{MayerMacro}
\bibinfo{author}{\bibfnamefont{C.}~\bibnamefont{Mayer}},
  \bibinfo{author}{\bibfnamefont{F.}~\bibnamefont{Sciortino}},
  \bibinfo{author}{\bibfnamefont{C.~N.} \bibnamefont{Likos}},
  \bibinfo{author}{\bibfnamefont{P.}~\bibnamefont{Tartaglia}},
  \bibinfo{author}{\bibfnamefont{H.}~\bibnamefont{Loewen}}, \bibnamefont{and}
  \bibinfo{author}{\bibfnamefont{E.}~\bibnamefont{Zaccarelli}},
  \bibinfo{journal}{Macromolecules} \textbf{\bibinfo{volume}{42}},
  \bibinfo{pages}{423} (\bibinfo{year}{2009}).

\bibitem[{\citenamefont{Young and Alder}(1977)}]{Young77}
\bibinfo{author}{\bibfnamefont{D.~A.} \bibnamefont{Young}} \bibnamefont{and}
  \bibinfo{author}{\bibfnamefont{B.~J.} \bibnamefont{Alder}},
  \bibinfo{journal}{Phys. Rev. Lett.} \textbf{\bibinfo{volume}{38}},
  \bibinfo{pages}{1213} (\bibinfo{year}{1977}).

\bibitem[{\citenamefont{Duran}(1999)}]{duran}
\bibinfo{author}{\bibfnamefont{J.}~\bibnamefont{Duran}},
  \emph{\bibinfo{title}{Sands and Powders and Grains: An Introduction to the
  Physics of Granular Materials}} (\bibinfo{publisher}{Springer, New York},
  \bibinfo{year}{1999}).

\bibitem[{\citenamefont{Fomin and Tsiok}(2013)}]{FominSilica}
\bibinfo{author}{\bibfnamefont{Y.~D.} \bibnamefont{Fomin}} \bibnamefont{and}
  \bibinfo{author}{\bibfnamefont{E.~N.} \bibnamefont{Tsiok}},
  \bibinfo{journal}{Phys. Rev. E} \textbf{\bibinfo{volume}{87}},
  \bibinfo{pages}{042122} (\bibinfo{year}{2013}).

\bibitem[{\citenamefont{Jagla}(1999)}]{Jagla}
\bibinfo{author}{\bibfnamefont{E.~A.} \bibnamefont{Jagla}},
  \bibinfo{journal}{J. Chem. Phys.} \textbf{\bibinfo{volume}{111}},
  \bibinfo{pages}{8980} (\bibinfo{year}{1999}).

\bibitem[{\citenamefont{de~Oliveira et~al.}(2008)\citenamefont{de~Oliveira,
  Netz, and C.}}]{Oliveira}
\bibinfo{author}{\bibfnamefont{A.}~\bibnamefont{de~Oliveira}},
  \bibinfo{author}{\bibfnamefont{P.~A.} \bibnamefont{Netz}}, \bibnamefont{and}
  \bibinfo{author}{\bibfnamefont{B.~M.} \bibnamefont{C.}},
  \bibinfo{journal}{Eur. Phys. J. B} \textbf{\bibinfo{volume}{64}},
  \bibinfo{pages}{481} (\bibinfo{year}{2008}).

\bibitem[{\citenamefont{{Ziherl} and {Kamien}}(2000)}]{Ziherl00}
\bibinfo{author}{\bibfnamefont{P.}~\bibnamefont{{Ziherl}}} \bibnamefont{and}
  \bibinfo{author}{\bibfnamefont{R.~D.} \bibnamefont{{Kamien}}},
  \bibinfo{journal}{Phys. Rev. Lett.} \textbf{\bibinfo{volume}{85}},
  \bibinfo{pages}{3528} (\bibinfo{year}{2000}).

\bibitem[{\citenamefont{{Malescio} and {Pellicane}}(2003)}]{Pellicane03}
\bibinfo{author}{\bibfnamefont{G.}~\bibnamefont{{Malescio}}} \bibnamefont{and}
  \bibinfo{author}{\bibfnamefont{G.}~\bibnamefont{{Pellicane}}},
  \bibinfo{journal}{Nat. Mater.} \textbf{\bibinfo{volume}{2}},
  \bibinfo{pages}{97} (\bibinfo{year}{2003}).

\bibitem[{\citenamefont{Osterman et~al.}(2007)\citenamefont{Osterman, Babic,
  Poberaj, Dobnikar, and Ziherl}}]{Osterman07}
\bibinfo{author}{\bibfnamefont{N.}~\bibnamefont{Osterman}},
  \bibinfo{author}{\bibfnamefont{D.}~\bibnamefont{Babic}},
  \bibinfo{author}{\bibfnamefont{I.}~\bibnamefont{Poberaj}},
  \bibinfo{author}{\bibfnamefont{J.}~\bibnamefont{Dobnikar}}, \bibnamefont{and}
  \bibinfo{author}{\bibfnamefont{P.}~\bibnamefont{Ziherl}},
  \bibinfo{journal}{Phys. Rev. Lett.} \textbf{\bibinfo{volume}{99}},
  \bibinfo{pages}{248301} (\bibinfo{year}{2007}).

\bibitem[{\citenamefont{Dotera et~al.}(2014)\citenamefont{Dotera, Oshiro, and
  Ziherl}}]{ZiherlNat2014}
\bibinfo{author}{\bibfnamefont{T.}~\bibnamefont{Dotera}},
  \bibinfo{author}{\bibfnamefont{T.}~\bibnamefont{Oshiro}}, \bibnamefont{and}
  \bibinfo{author}{\bibfnamefont{P.}~\bibnamefont{Ziherl}},
  \bibinfo{journal}{Nature} \textbf{\bibinfo{volume}{506}},
  \bibinfo{pages}{208} (\bibinfo{year}{2014}).

\bibitem[{\citenamefont{{Sperl} et~al.}(2010)\citenamefont{{Sperl},
  {Zaccarelli}, {Sciortino}, {Kumar}, and {Stanley}}}]{Sperl2010}
\bibinfo{author}{\bibfnamefont{M.}~\bibnamefont{{Sperl}}},
  \bibinfo{author}{\bibfnamefont{E.}~\bibnamefont{{Zaccarelli}}},
  \bibinfo{author}{\bibfnamefont{F.}~\bibnamefont{{Sciortino}}},
  \bibinfo{author}{\bibfnamefont{P.}~\bibnamefont{{Kumar}}}, \bibnamefont{and}
  \bibinfo{author}{\bibfnamefont{H.~E.} \bibnamefont{{Stanley}}},
  \bibinfo{journal}{Phys. Rev. Lett.} \textbf{\bibinfo{volume}{104}},
  \bibinfo{pages}{145701} (\bibinfo{year}{2010}).

\bibitem[{\citenamefont{Sellitto}(2013)}]{SellittoJCP2013}
\bibinfo{author}{\bibfnamefont{M.}~\bibnamefont{Sellitto}},
  \bibinfo{journal}{J. Chem Phys.} \textbf{\bibinfo{volume}{138}},
  \bibinfo{pages}{224507} (\bibinfo{year}{2013}).

\bibitem[{\citenamefont{Das et~al.}(2013)\citenamefont{Das, Gnan, Sciortino,
  and Zaccarelli}}]{Gayatri}
\bibinfo{author}{\bibfnamefont{G.}~\bibnamefont{Das}},
  \bibinfo{author}{\bibfnamefont{N.}~\bibnamefont{Gnan}},
  \bibinfo{author}{\bibfnamefont{F.}~\bibnamefont{Sciortino}},
  \bibnamefont{and}
  \bibinfo{author}{\bibfnamefont{E.}~\bibnamefont{Zaccarelli}},
  \bibinfo{journal}{J. Chem. Phys.} \textbf{\bibinfo{volume}{138}},
  \bibinfo{pages}{134501} (\bibinfo{year}{2013}).

\end{thebibliography}

\begin{thebibliography}{99}
\bibitem{Das} G. Das, N. Gnan, F. Sciortino and E. Zaccarelli, J. Chem. 
Phys. \textbf{138}, 134501 (2013).
\bibitem{Franosch1997}T. Franosch, M. Fuchs, W. G{\"o}tze, M. R. Mayr, and 
A. P. Singh, Phys. Rev. E \textbf{55} 7153 (1997).
\bibitem{Goetze2002}W. G{\"o}tze and M. Sperl, Phys. Rev. E \textbf{66}, 
011405 (2002). 
\bibitem{Sci03a} F. Sciortino, P. Tartaglia, E. Zaccarelli, Phys. Rev. Lett. \textbf{91}, 268301 (2003).
\bibitem{SSSPRL}M. Sperl, E. Zaccarelli, F. Sciortino, P. Kumar,
H. E. Stanley, Phys. Rev. Lett. {\bf 104}, 145701 (2010).
\end{thebibliography}

\newpage
\onecolumngrid
\setcounter{figure}{0}
\section*{Mapping of MCT fluid-glass line on simulation data}

\begin{figure}[h]
\begin{center}
\includegraphics[width=0.7\textwidth]{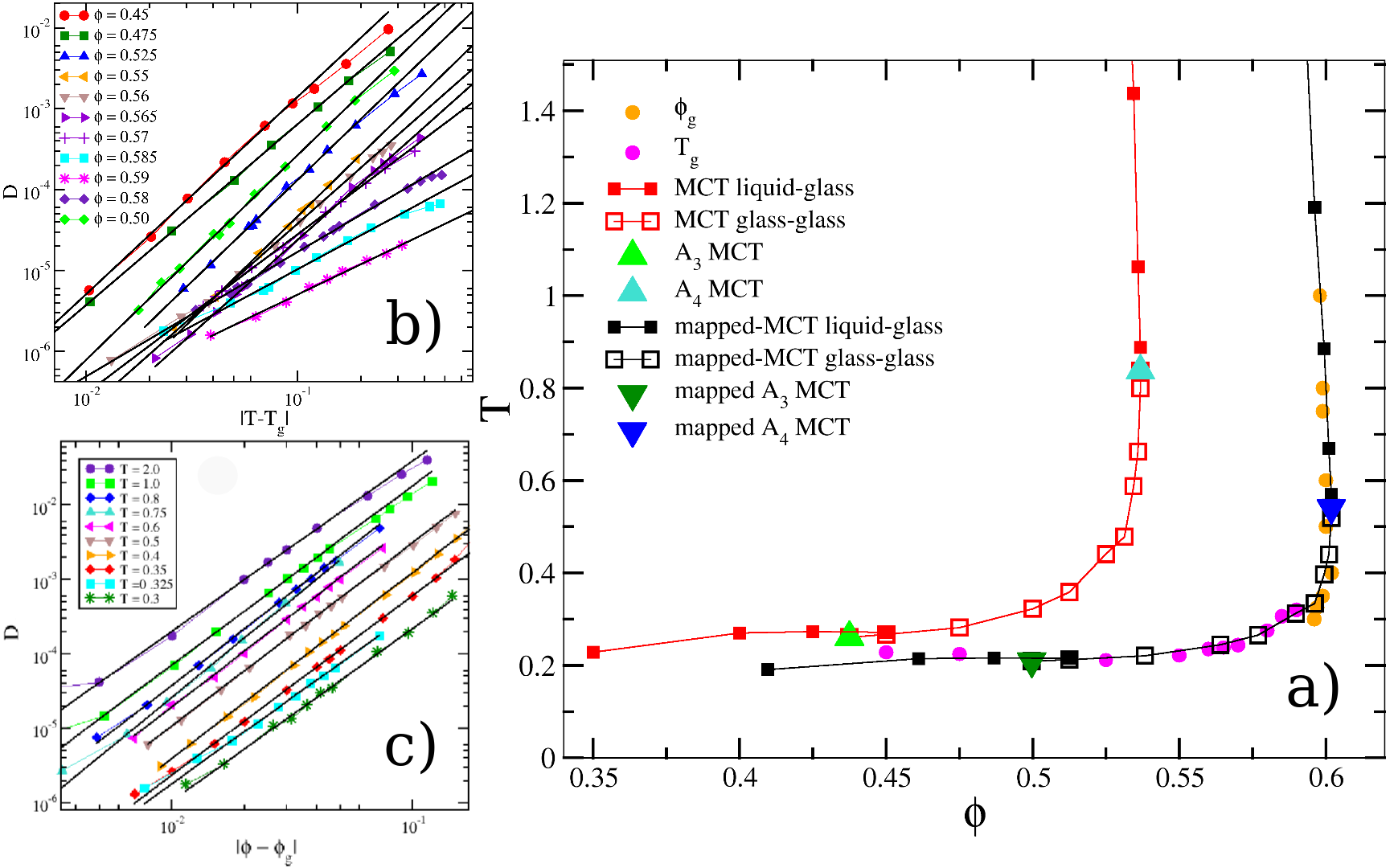}
\end{center}
\caption{ (a) Fluid-glass and glass-glass transition line from MCT (red curve with open and filled symbols) obtained using as input the static structure 
factor evaluated from simulations.The magenta and orange filled circles are glass transition temperatures $T_g$ and packing 
fractions $\phi_g$ extracted from a power-law fitting of the diffusivity $D$ along isotherms and isochores respectively.
A bilinear transformation in $\phi$ and $T$ allows us to superimpose the theoretical data on the simulation results (black curve with open and filled symbols). (b) Diffusivity of species $A$, $D$ as a functon of $T-T_g$. (c)$D$ as a function of $\phi-\phi_g$.}\label{fig:simulation-mct}
\end{figure}

To investigate the presence of two $A_4$ points, we solve the MCT for the monodisperse SS system finding the presence of the two
higher-order singularities within the Rogers-Young (RY) closure. These predictions have been confirmed by repeating the MCT 
calculations for our binary mixture, using as input of the theory the structure factors $S(q)$ evaluated in simulations, 
thus avoiding to rely on a specific closure of the Ornstein-Zernike equation. From these calculations we estimate 
$\Delta^{*(1)}_{MCT}\simeq 0.17$ and $\Delta^{*(2)}_{MCT}\simeq 0.20$. As already observed in previous works, the MCT 
gives a qualitative, but not quantitative, estimation of the fluid-glass line and of higher-order singularities. 
Hence, to correctly locate the $A_4$ we need to map the theoretical fluid-glass line on that extrapolated from simulations. 
The latter can be achieved from a well established procedure:  we perform simulations for several state points at 
intermediate and low diffusivities in order to extract the diffusion coefficient $D$ of the species $A$ from the 
long-time limit of the mean-square displacement. On approaching the ideal glass line MCT predicts that $D$ tends to zero 
following the laws $D\sim\mid\phi-\phi_g(T)\mid^{\gamma(T)}$ and $D\sim\mid T-T_g(\phi)\mid^{\gamma(\phi)}$, where 
$\phi_g$ and $T_g$ are, respectively, the value of $\phi$ and of $T$ at the fluid-glass transition and $\gamma$ is an 
exponent that is determined by the theory. The power-law fits for $D$ as a function of $|T-T_g|$ and $|\phi-\phi_g|$ are 
shown in Fig.~\ref{fig:simulation-mct} (b) and (c). The power-law behavior of $D$ allows us to trace the locus of points 
in the ($\phi$,$T$) phase diagram for which $D=0$ as shown in Fig.~\ref{fig:simulation-mct}(b).

We then perform a bilinear transformation $T\rightarrow 0.575 T +0.059$, $\phi \rightarrow 1.03 \phi +0.049$ to 
superimpose the theoretical and the simulation line as reported in Fig.~\ref{fig:simulation-mct} (a), finding that the 
first $A_4$ is located at $(\phi^*=0.6,T^*=0.54)$. As stated above, the presence of an $A_4$ is signalled by a pure 
logarithmic behaviour of the density correlator at a given wave vector.

\begin{figure}[h]
\begin{center}
\includegraphics[width=0.4\textwidth]{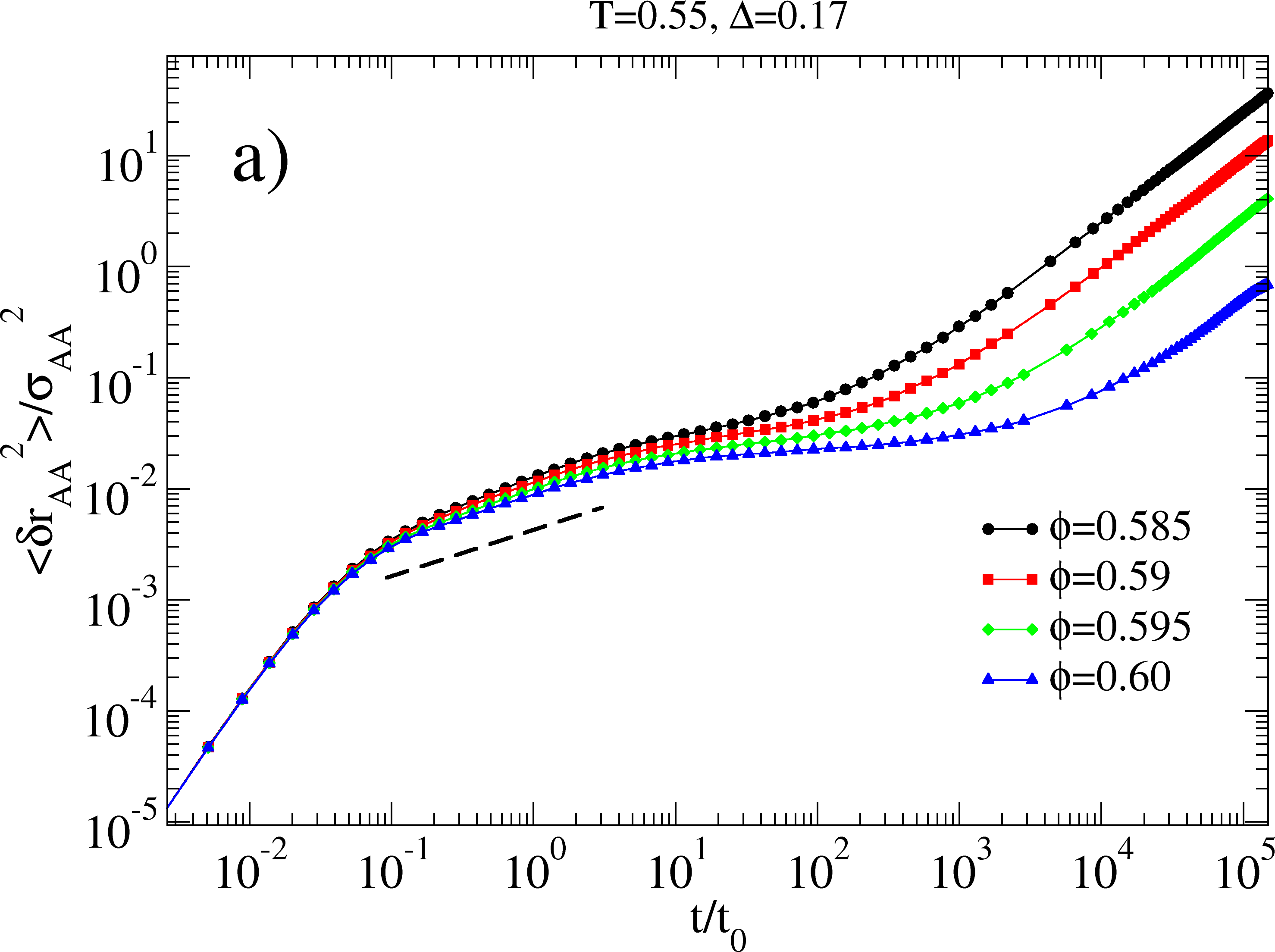}~%\\\vspace{0.5cm}
\includegraphics[width=0.4\textwidth]{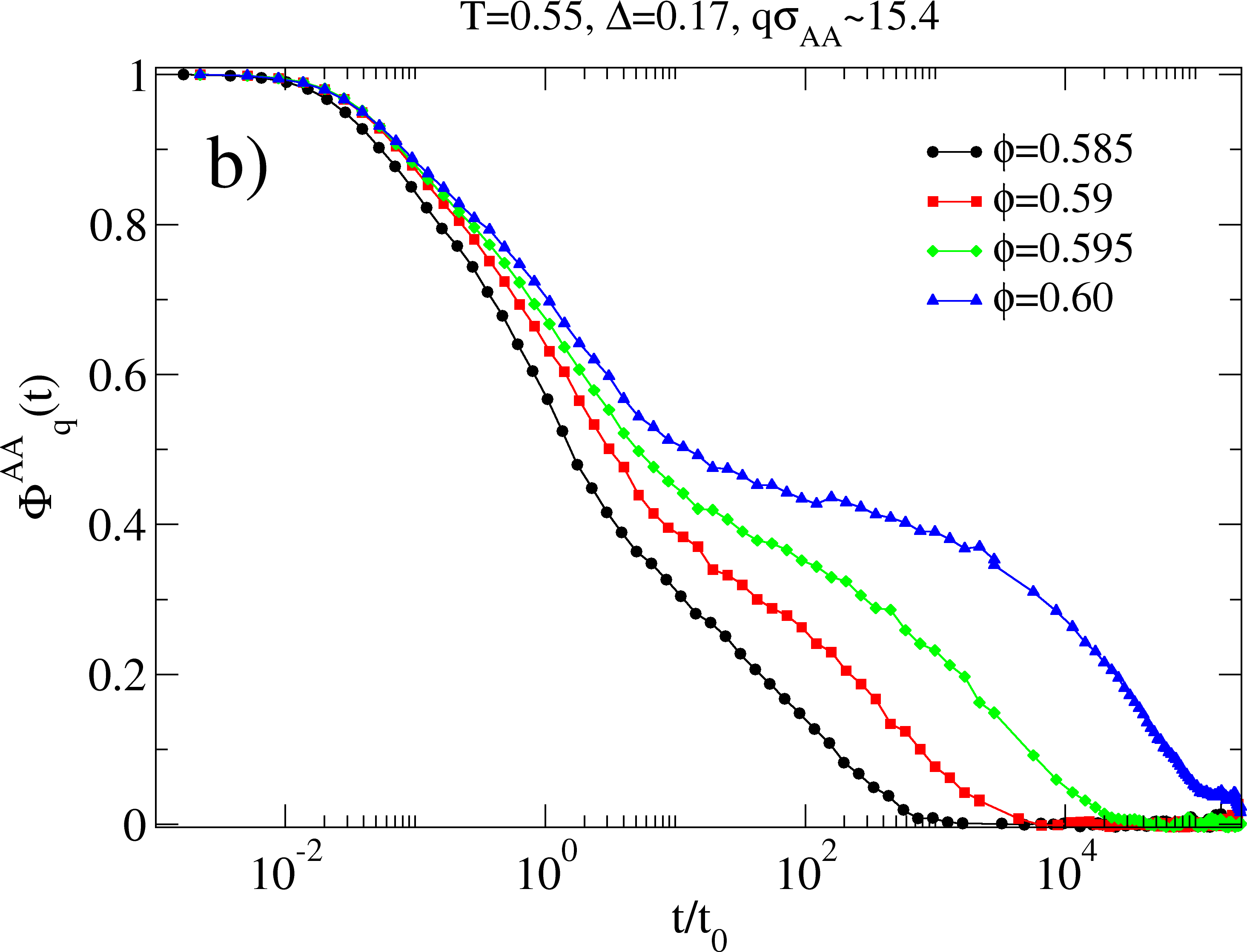}
\caption{a) MSD for the A particles at $T=0.55$, $\Delta=0.17$ and different $\phi$. At the state point corresponding to the $A_4$ obtained by mapping  the MCT fluid-glass line on to the simulation data, only a very weak influence of a higher order singularity can be observed. This is underlined by the dashed line which is a guide to the eye. b) density autocorrelation function $\Phi_q^{AA}(t)$ for the same state points of (a), for $q\sigma_{AA}\sim15.4$ (the wave vector slightly changes depending on $\phi$). }
\label{fig:MSD_DELTA0_17}
\end{center}
\end{figure}

We have thus evaluated the density autocorrelation function of the A species $\Phi_q^{AA}(t)$ at different wave vectors 
$q$ in a region of $T$ and $\phi$ close to the estimated $A_4$, together with the mean square displacement (MSD) 
$\langle\delta r_{AA}^2\rangle$ as shown in Fig. \ref{fig:MSD_DELTA0_17} (a) and (b).

\begin{figure}[h]
\begin{center}
\includegraphics[width=0.4\textwidth]{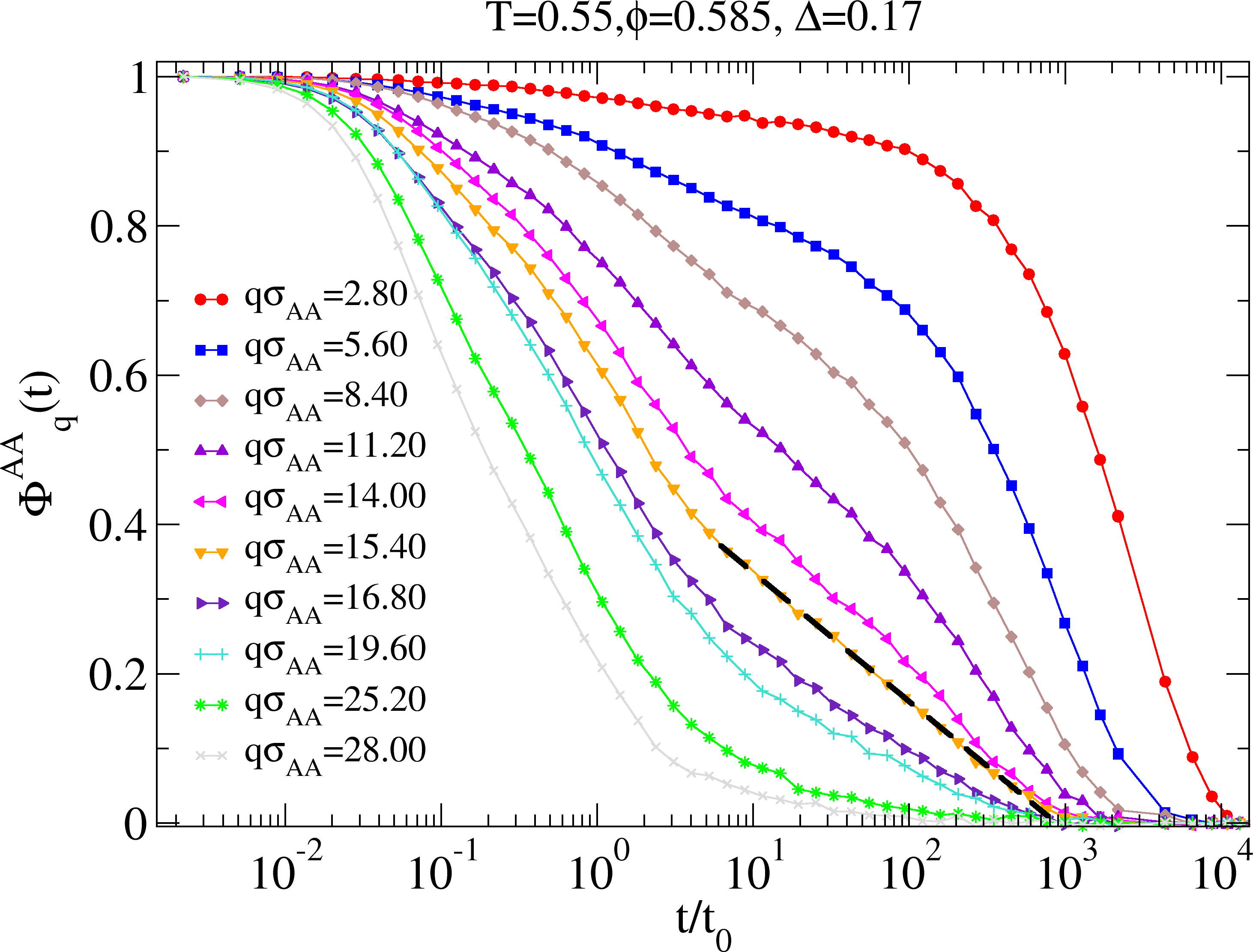}
\caption{density autocorrelation function $\Phi_q^{AA}(t)$ of the A species at $T=0.55$,$\phi=0.585$,$\Delta=0.17$ for different wave vectors. At such $\phi$, for $q\sigma_{AA}=15.4$, the last part of the decay can be fitted with a pure logarithm, showing the presence of a higher-order singularity which is far from the given state point.}
\label{fig:FqA3}
\end{center}
\end{figure}

\noindent Notice that, at the expected state point we do not find any signature of an $A_4$ singularity. Instead the influence of the fluid-glass line controls the evolution of the  two dynamic observables, thus hiding the presence of higher order singularities. In the present case, we observe the presence of an $A_3$ point, which lays in the glass region, but which can still influence the fluid states. As already found previously \cite{Das}, the $A_3$ contribution to the dynamics can be observed only when the effect of the fluid glass-line on the fluid states is small. This is clearly shown in Fig. \ref{fig:MSD_DELTA0_17} where at $\phi=0.585$ a logarithmic decay is found in $\Phi_q^{AA}(t)$, although for only two decades, and it gets progressively lost on increasing $\phi$, similarly to what observed for $\Delta=0.15$\cite{Das}.
For such state point we have also evaluated $\Phi_q^{AA}(t)$ for different wave vectors $q\sigma_{AA}$ to show the typical concave-to-convex behaviour as illustrated in Fig.\ref{fig:FqA3}.  Since we find only $A_3$ features, we must increase $\Delta$ in order to find the $A_4$ singularity (following MCT predictions in Fig.1 of the manuscript). Indeed, we find such singularity for $\Delta\simeq 0.21$ as discussed in the manuscript.

\section*{Energy and structure along Iso-Dynamics loci}

\begin{figure}[h]
\begin{center}
\includegraphics[width=0.4\textwidth]{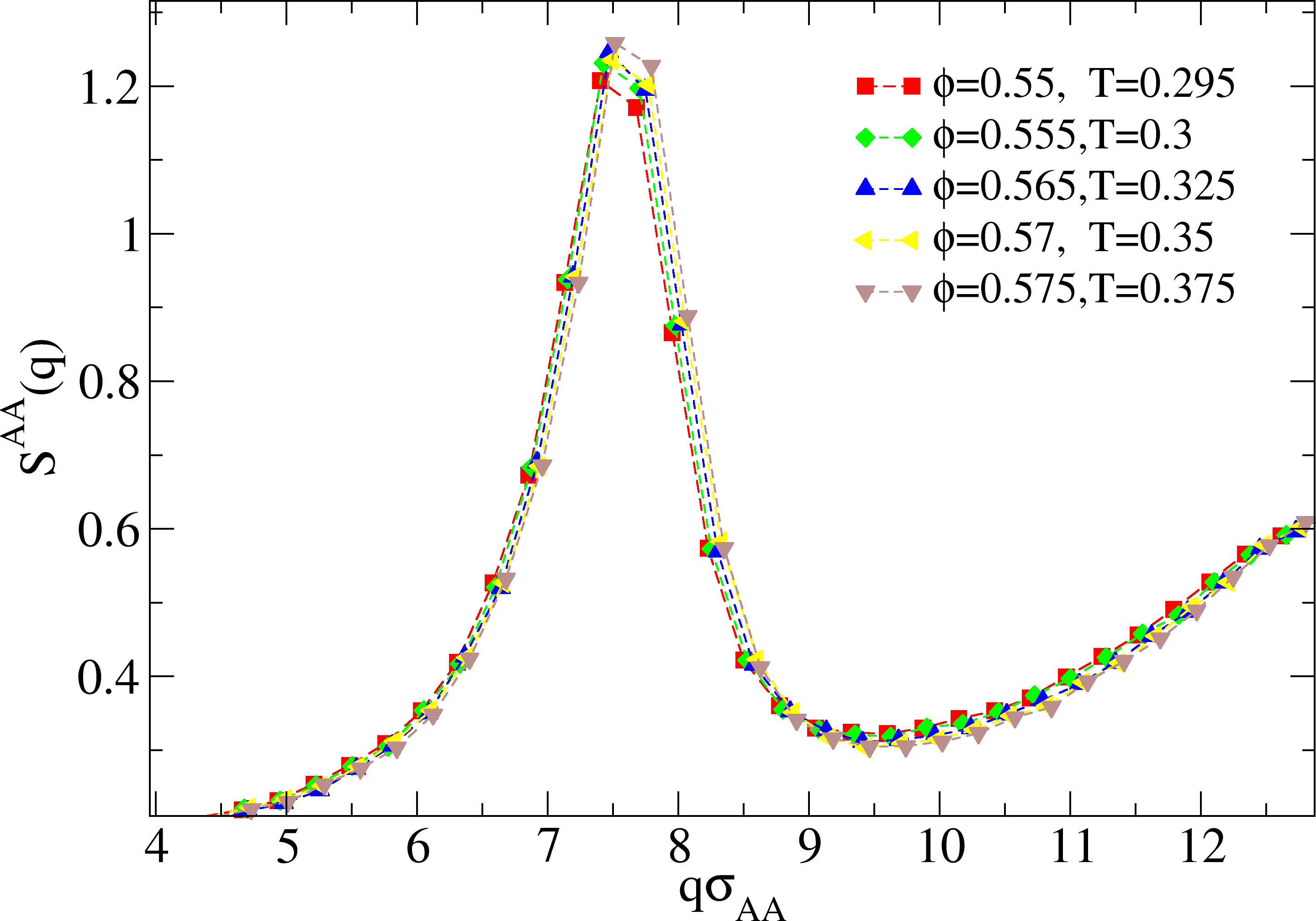}

\caption{Static structure factor for invariant point along the iso-$D/D_0$ with  $D/D_{0}=5.5\times10^{-5}$ (squares).}
\label{fig:sq-invariantline}
\end{center}
\end{figure}

Here we show that while dynamics is invariant, thermodynamic and structural properties are not. 
Fig.~\ref{fig:sq-invariantline} shows the structure factor $S(q)$ of a set of invariant points belonging to the 
$iso-D/D_0$ curve with $D/D_{0}=5.5\times10^{-5}$. Since the iso-dynamics points are very close to each other (there is 
only a difference of $4\%$ in $\phi$ for the outermost points identified), $S(q)$ is very similar for all the state 
points, but we do observe a trend, e.g. in the growth of the first peak, which seems genuine within numerical resolution.

\begin{figure}[h]
\begin{center}
\includegraphics[width=0.4\textwidth]{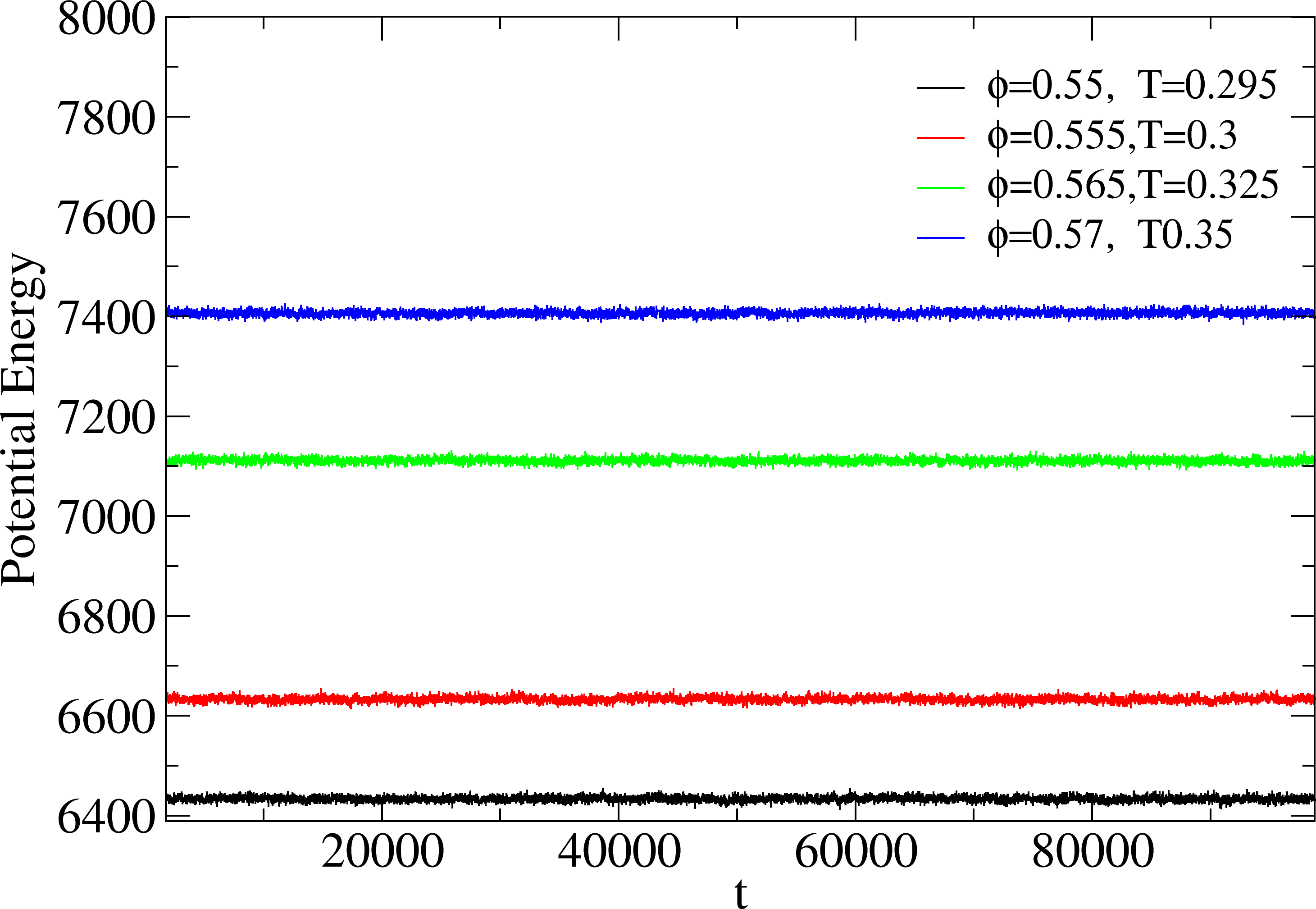}
\caption{Total potential energy for invariant point along the iso-$D/D_0$ with  $D/D_{0}=5.5\times10^{-5}$ (squares).}
\label{fig:energies_invariantlines}
\end{center}
\end{figure}

A stronger difference is observed by looking at the potential energy $U$ along an iso-dynamics line. Since $U=\int dr 
g(r) u(r_{ij})$ (where, $u(r_{ij})$ is the pair potential and $g(r)$ is the radial distribution function), different $U$ 
values imply a different structure for each point along the iso-dynamics line. Fig.\ref{fig:energies_invariantlines} 
shows the total potential energy for invariant points having $D/D_{0}=5.5\times10^{-5}$.

\section*{Iso-dynamics within MCT}

\begin{figure}[h]
\begin{center}
\includegraphics[width=0.4\textwidth]{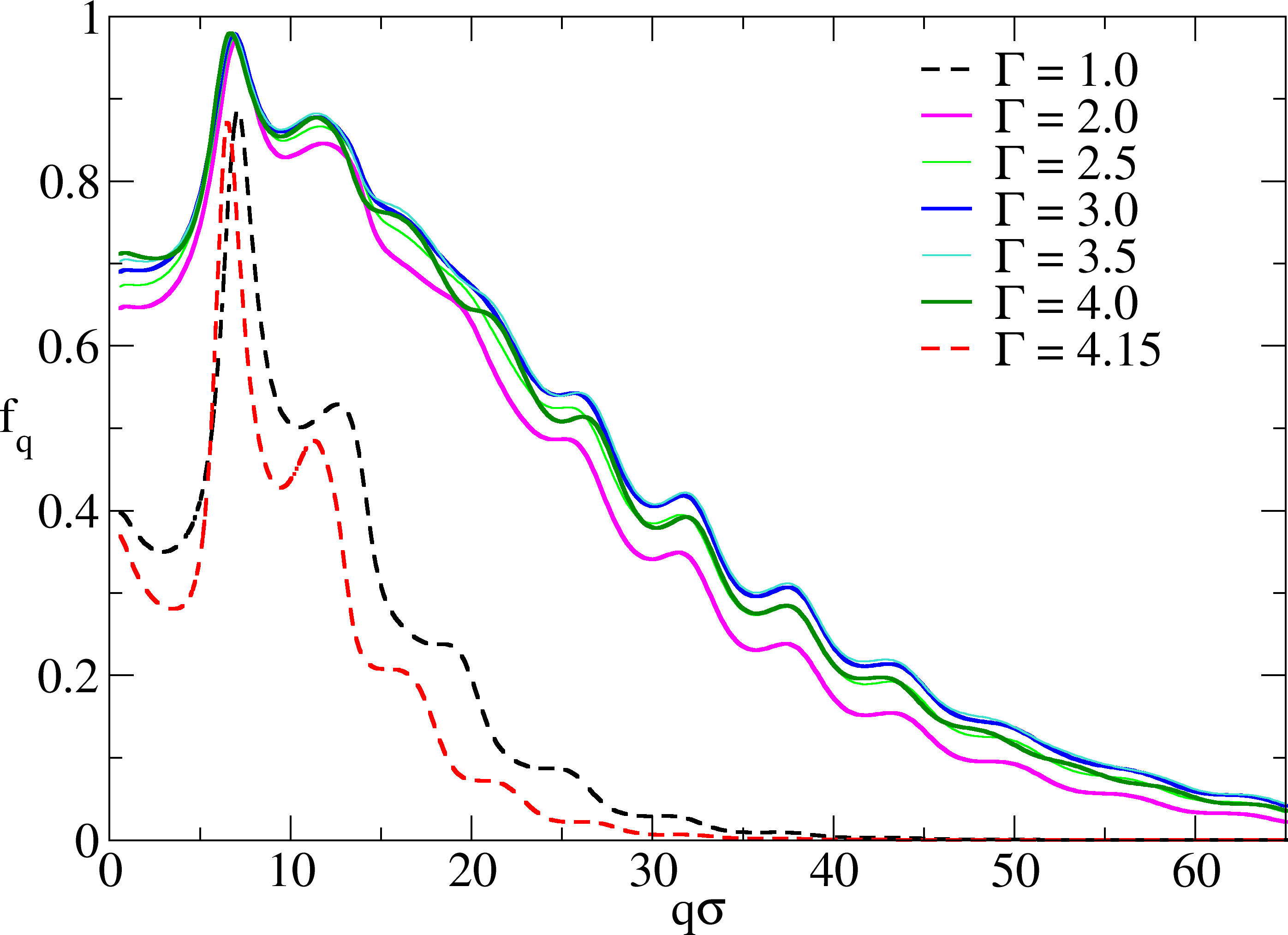}
\caption{Glass-form factors $f_q$ along the transition line for $\Delta$ = 
0.17 and various values for the inverse temperature $\Gamma = 1/T$ 
calculated from MCT within RY, cf. \cite{SSSPRL}. The dashed curves 
represent transition points close to the higher-order singularities but 
\textit{outside} the region of invariant dynamics. The full curves show 
the results for points spaced equally along the line \textit{inside} the 
region of invariant dynamics.  
}
\label{fig:MCTSSSRYfq}
\end{center}
\end{figure}

While there is no general perturbative description for a classical fluid, 
it is the putative existence of MCT singularities that allows for such an 
analytical expansion. For both fluid-glass as well as higher-order 
singularities, such expansions involve the wave-vector dependent 
coefficients $f_q$, $H^{(1)}_q$, $H^{(2)}_q$
%$h_q$, $K_q$, $\dots$ 
\cite{Franosch1997,Goetze2002,Sci03a}. 
While these coefficients experience a regular dependence on the static 
structure, i.e. the external control parameters, they vary discontinuously 
at crossings of glass-transition lines which occur generically in the 
vicinity of higher-order singularities. As shown in \cite{SSSPRL}, the 
higher-order singularities originate from a delicate balance of inner and 
outer shell of the potential: In between the higher-order singularities, 
the two contact values in the pair distribution functions $g(r)$ cause a 
beating in the static structure factor which gives rise to a strong enough 
additional contribution in the MCT vertex to trigger a line of glass-glass 
transition with endpoint singularities. It is important to notice that the 
two endpoints represent symmetric extreme representatives of such 
glass-glass transition points.

\begin{figure}[h]
\begin{center}
\includegraphics[width=0.4\textwidth]{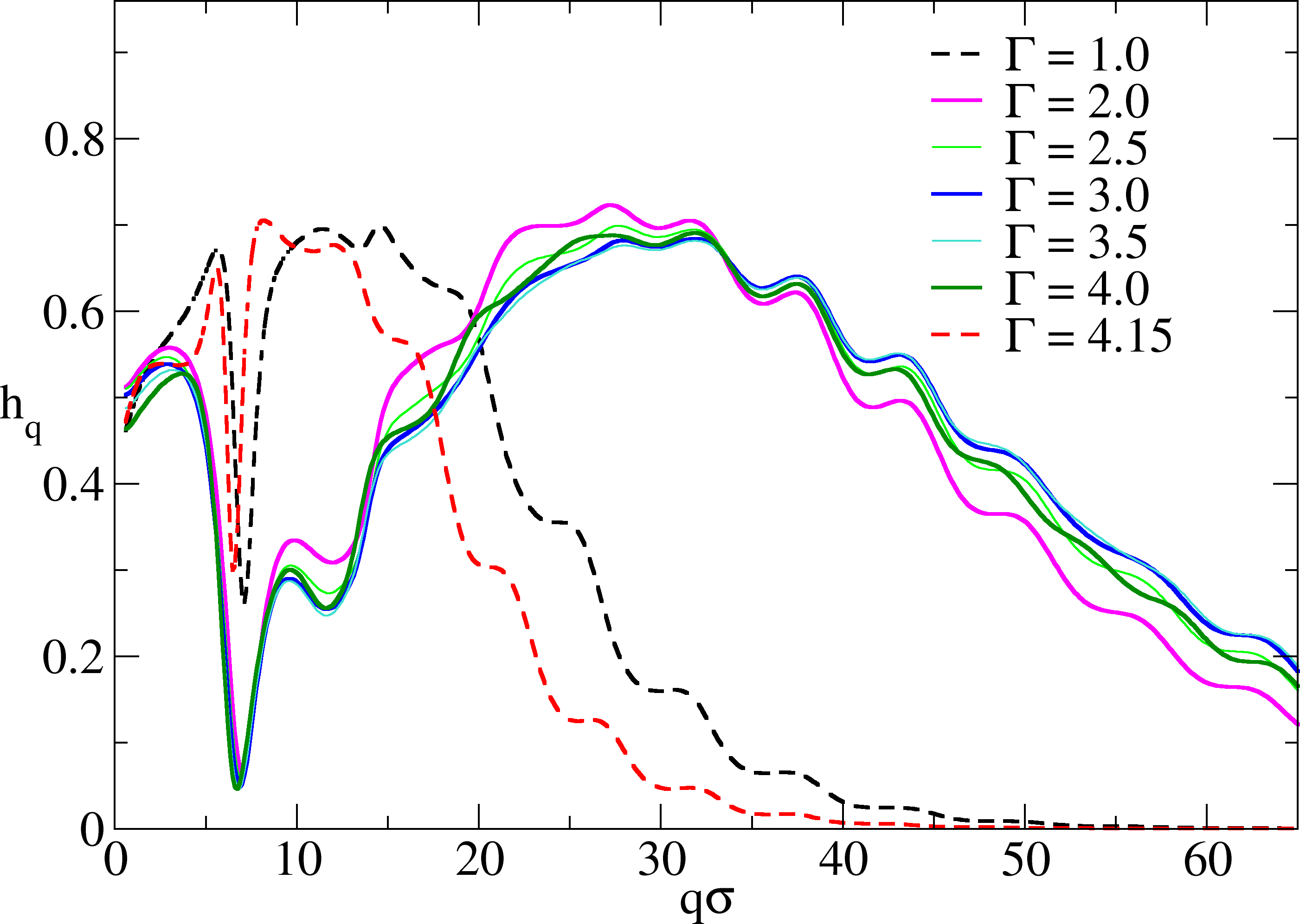}
\caption{Critical amplitudes $h_q$ along the transition line for $\Delta$ 
= 0.17 and various values for the inverse temperature $\Gamma = 1/T$ 
calculated from MCT within RY, cf. \cite{SSSPRL}. Line styles are 
identical to Fig.~\ref{fig:MCTSSSRYfq}.
}
\label{fig:MCTSSSRYhq}
\end{center}
\end{figure}

In between the endpoints but not outside, the glass-transition properties 
are generically very similar as shown for $f_q$ in 
Fig.~\ref{fig:MCTSSSRYfq}. Glass-form factors $f_q$ are shown for several 
transition points in between the higher-order singularities as full curves 
together with two cases as dashed lines just outside the region. While the 
distributions outside vary from each other in a regular fashion, the 
distributions inside show very little variation. The situation is similar 
for the critical amplitude $h_q$ in Fig.~\ref{fig:MCTSSSRYhq}, and also in 
the same way behave values for the localization lengths, tagged-particle 
correlators, and other amplitudes.

The iso-dynamics lines do not extend all the way towards the higher-order 
singularities since at the higher-order singularities the dynamics is 
modified significantly from its well-known behavior by the critical 
exponents changing rapidly over a very narrow region of control parameters 
\cite{Goetze2002}. In between, the critical exponents $a$ and $b$ (the MCT 
exponent parameter $\lambda$), have a broad maximum (minimum). Hence, a 
generic reason for the iso-dynamics lines can be identified in the presence 
of two symmetric MCT endpoint singularities: The short-time dynamics is 
unaffected by the presence of glass transitions and is therefore similar 
if the variation of the statics structure factors is small. The dynamics 
affected by the glass transitions experience especially little variation 
in the region of invariant dynamics since the variation of the critical 
parameters are restricted as explained above.

\section*{Indications from the Static Structure}

\begin{figure}[h]
\begin{center}
\includegraphics[width=0.4\textwidth]{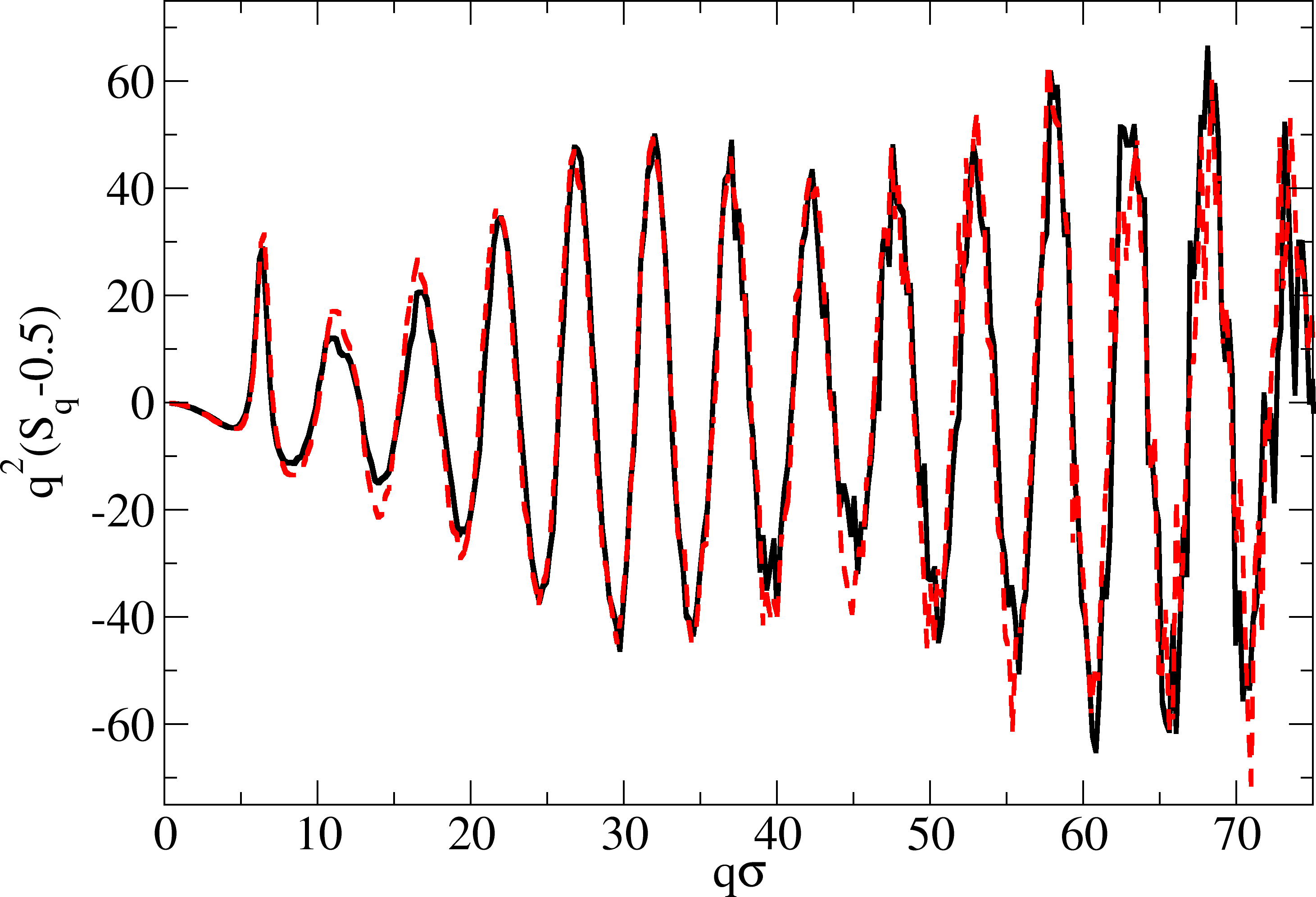}
\caption{Static structure factor for large wave vectors plotted as
$q^2(S_q-0.5)$ for $\Delta = 0.17$ and
$(\varphi, T) = (0.55, 0.295)$ as full and (0.58, 0.425) 
as dashed curve.
}
\label{fig:beating}
\end{center}
\end{figure}

Within the MCT calculations it could be shown that the higher-order singularities
originate from the competition of different wave-vector regimes in the static structure 
factor $S_q$ where a beating phenomenon reflects the influence of inner and outer shell  
\cite{SSSPRL}. To support the conclusions drawn above within the MCT picture we show in
Fig.~\ref{fig:beating} the static structure factor for the big particles: The contributions
above the regular decay of $1/q^2$ for $S_q$ towards its large-$q$ limit of 0.5 are clearly 
seen to exhibit beating with a frequency characteristic of the distance of the two shells, 
$2\pi/\Delta \approx 37$. This beating frequency is observed most clearly between $q\sigma= 15$
and 45 while mixing effects may interfere at yet larger wave vectors and obscure the visibility
of the beating. Finding the beating phenomenon in a region where also MCT effects regarding the
higher-order singularities are most prominent lends further support to the interpretation
of our data within the MCT predictions.

\end{document}